%% file: Manuscript.tex
\documentclass[prx, twocolumn, 10pt,
nobibnotes,superscriptaddress,
 amsmath,amssymb,
 aps,longbibliography
]{revtex4-1}
\usepackage[normalem]{ulem}
\usepackage{graphicx}
\usepackage{dcolumn}
\usepackage{bm}
\usepackage{hyperref} 
\usepackage{color}
\usepackage{amsfonts,latexsym}

\definecolor{mygreen}{rgb}{0.1, 0.6, 0.4}

\newcommand\redout{\bgroup\markoverwith
{\textcolor{blue}{\rule[.5ex]{2pt}{0.4pt}}}\ULon}

\begin{document}
\preprint{}

\title{Control of Stochastic and Induced Switching in Biophysical Networks}

\author{Daniel K. Wells}
\affiliation{Department of Engineering Sciences and Applied Mathematics, Northwestern University, Evanston, IL 60208, USA}
\affiliation{Northwestern Physical Sciences-Oncology Center, Northwestern University, Evanston, IL 60208, USA}

\author{William L. Kath}
\affiliation{Department of Engineering Sciences and Applied Mathematics, Northwestern University, Evanston, IL 60208, USA}
\affiliation{Northwestern Physical Sciences-Oncology Center, Northwestern University, Evanston, IL 60208, USA}
\affiliation{Northwestern Institute on Complex Systems, Northwestern University, Evanston, IL 60208, USA}

\author{Adilson E. Motter} \email{Electronic Address: motter@northwestern.edu}
\affiliation{Northwestern Physical Sciences-Oncology Center, Northwestern University, Evanston, IL 60208, USA}
\affiliation{Northwestern Institute on Complex Systems, Northwestern University, Evanston, IL 60208, USA}
\affiliation{Department of Physics and Astronomy, Northwestern University, Evanston IL, 60208, USA}


\begin{abstract}Noise caused by fluctuations at the molecular level is a fundamental part of intracellular processes. While the response of biological systems to noise has been studied extensively, there has been limited understanding of how to exploit it to induce a desired cell state. Here we present a scalable, quantitative method based on the Freidlin-Wentzell action to predict and control noise-induced switching between different states in genetic networks that, conveniently, can also control transitions between stable states in the absence of noise. We apply this methodology to models of cell differentiation and show how predicted manipulations of tunable factors can induce lineage changes, and further utilize it to identify new candidate strategies for cancer therapy in a cell death pathway model. This framework offers a systems approach to identifying the key factors for rationally manipulating biophysical dynamics, and should also find use in controlling other classes of noisy complex networks.\\

Subject Areas: Complex Systems, Biological Physics, Nonlinear Dynamics
\end{abstract}


\maketitle

\section{Introduction}
Cellular systems are not entirely deterministic, but are instead impacted 
by small, random fluctuations in the number and activity of molecules  of
intracellular species~\cite{Raj:2008ip,Arkin:1997dw}. Such fluctuations lead to macroscopic effects in a diverse array of processes. In differentiation, the resulting noise plays a central role in cell fate determination and can allow clonal populations of differentiating cells to achieve distinct final states~\cite{Balazsi:2011id,Chang:2008gu}. Noise can also produce spontaneous transitions, whereby it causes a system to switch from one stable state to another, often  producing a significant change of phenotype or function. Such stochastic state switching occurs, for example, in the \textit{lac} system, where rare, brief transcription events in the ``off'' state cause large bursts in LacY expression, which in turn can be amplified and stabilized by a positive feedback loop~\cite{Choi:2008ke}. Stochastically-induced transitions also underlie recent observations of spontaneous dedifferentiation in cancer cells~\cite{Gupta:2011fk,Chaffer:2011iv}, in which cancer stem cells arose de-novo from non-stem cell populations.

The response to noise and the overall behavior of many biophysical systems are determined by an underlying epigenetic landscape~\cite{WADDINGTON:1957wn}. In this landscape, the valleys represent the distinct achievable states of the system and the heights of the separating barriers determine their robustness to noise. A 
benefit of this representation is that bifurcation points---locations in the parameter space at which one or more stable states suddenly cease to exist---correspond precisely to the points
where  one or more of such barriers first reach zero height as parameters change.
This landscape thus incorporates two distinct features of a state, namely its robustness to noise and its
deterministic stability, into one: {\it the less robust a state is to noise, the closer it is to being eliminated through a bifurcation, and vice-versa}. 

The landscape representation 
has been given a quantitative foundation as the quasipotential of the  
deterministic component of the
system dynamics~\cite{Huang:2005fv} and has been explored in experiments---e.g.,\ to show how two  parameters in the yeast galactose signaling network, the concentrations of galactose and intracellular Gal80p, 
can alter the rates of stochastic switching in this bistable circuit~\cite{Acar:2005ds}.  
Despite these advances, researchers' ability to control this landscape in order to induce prespecified biological outcomes
has been generally limited to  at most two 
parameters~\cite{Kiga, Gore}, and no general method currently exists to systematically tune transitions between stable states and/or eliminate undesired states altogether.
The possibility of such control would offer clear opportunities. 
For example, under the widely supported stochastic model for induced pluripotent stem cell generation~\cite{Hanna:2009ix}, a majority of cells have the possibility of being reprogrammed, even though 
existing technologies have achieved substantially smaller yields~\cite{Robinton:2012}. 
The ability to control the response to noise of 
differentiated and stem-cell states (e.g., inhibiting transitions to the first and promoting transitions to the second) could lead to
enhanced procedures to create induced pluripotent stem cells. Similarly, in the context of the ``cancer attractor'' hypothesis~\cite{Huang:2009kj,Creixell2012}, in which normal and cancer cells correspond to distinct co-existing stable states,  identifying interventions that destabilize, or eliminate, the cancerous state could lead to new therapeutic strategies. 

\begin{figure*}[t!]
\centerline{\includegraphics[width = 0.70\textwidth]{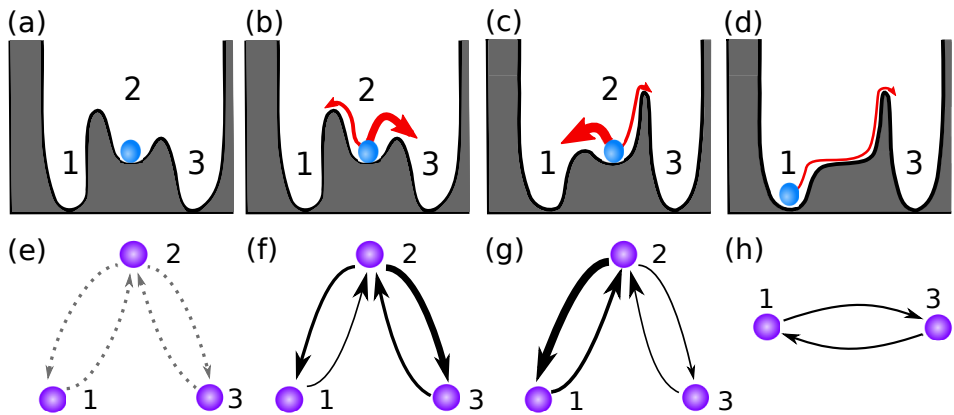}}
\caption{{ Control of the response to noise illustrated for a multi-well quasipotential.}
 (a) Without noise the state of the system is fixed and will not change over time. 
 (b) In the presence of noise the state of the system wanders within the attraction basin and will be eventually ejected into the attraction basin of another stable state. For small noise such transitions are exponentially more likely to occur through the lowest barrier, even though other transitions are also  possible in principle. 
 (c) Using OLAC  on a set of tunable parameters, we can alter the quasipotential to produce a desired response to noise, in this case tailored to increase transitions from state $2$ to state $1$ and to reduce transitions to state $3$. 
 (d) If desired, OLAC can also find combinations of tunable parameters that can alter the quasipotential in order to eliminate a stable state through a bifurcation; in this
 illustration state $2$ is eliminated, in favor of state $1$.
 (e-h) NEST for each of the quasipotentials in (a-d), respectively, where the nodes represent stable states and  the continuous edges represent transition rates. A wider continuous edge indicates a higher transition rate. The dotted edges in (e) indicate transitions that could occur in the presence of noise.
}
\label{fig:0}
\end{figure*}

In this paper 
we  propose a broadly applicable method, here termed \textit{optimal least action control} (OLAC), that can predict and control the dynamical behavior and response to noise in a wide class of biophysical networks.
As schematically illustrated in Fig.~\ref{fig:0}, the essence of our approach is that to control a biophysical 
system it is sufficient to identify interventions---e.g.,  changes to gene expression, protein levels, or interaction rates---that can reshape the topography of the underlying quasipotential in a desired way. This approach ultimately leads to a  
{\it network of state transitions} (NEST)
describing the transitions between stable states and that can be controlled by changing the heights of the separating barriers 
without changing 
any quality of the noise. 
For a given system, this is achieved by determining the minimum action paths---those followed by the most likely noise-induced transition trajectories---and the corresponding transition rates between all pairs of stable states, and then optimizing these transition rates for a desired outcome. Furthermore,  
this general foundation in a physical least-action principle allows OLAC to be applied broadly to many other complex networks as well. In particular, while we 
focus our application of OLAC to biophysical networks, applications to other networks where noise and multistability play important roles, including power-grid networks~\cite{MotterNatPhys}, polymer networks~\cite{MacKintoshMetastable} and food-web networks~\cite{Motter2}, among others, are 
immediate 
within the formulation we establish here.

We apply OLAC to several 
gene network models and illustrate how this method can be used to make biologically realizable reprogramming predictions.  In the limit of zero noise intensity,  
OLAC automatically identifies bifurcations that eliminate undesirable states and induce purely deterministic transitions to the desired ones. 
The significance of the latter
is demonstrated by considering a third application, to eliminate cancerous states in a cell death network model,
which concerns a time scale for which stochastic switches can be neglected.
 As illustrated in these examples, the NEST  is a powerful yet simple representation that captures the essence of the state switching dynamics and can inform counterintuitive results---e.g., 
 the possibility of transitions through intermediate stable states when direct transitions are essentially impossible (a behavior observed even in high dimensions, where indirect transitions generally require longer paths). The method proposed here is easily implemetable and the computational effort scales 
linearly in the number of control parameters and
the dimension of the state space,
allowing our approach to be applied to large networks and high-dimensional systems in general.

\section{Theory}
\subsection{Transition Rates for Small Noise} 
 We consider 
 biophysical networks
 whose deterministic components are described by non-linear differential equations of the form $d\vec{X}/{dt} = \vec{F}(\vec{X};\Omega)$, where $\vec{X}$ is a vector representing the activity of the relevant biological factors,  $\vec{F}$ is the function representing the rates of change  of these factors, and  $\Omega$ is a set of tunable parameters, which we show can be manipulated to drive cellular processes in advantageous directions.  
We focus on the most prevalent case of systems with two or more stable states and, although our approach is general, for concreteness we first assume that these states are time-independent.
Time-independent stable states  correspond to fixed points $\vec{X}^{*}=\vec{X}^{*}(\Omega)$, defined by $\vec{F}(\vec{X}^{*};\Omega) =\vec{0}$, towards which neighboring trajectories converge over time.  
The set of all  
stable states 
represent the possible 
long-term 
behaviors of the deterministic 
system.

Stochasticity is modeled here as additive Gaussian white noise, 
\begin{equation}
\label{eqn:0a}
d \vec{X} = \vec{F}(\vec{X};\Omega) dt + \sqrt{\varepsilon}\  d\vec{W},
\end{equation} where $\varepsilon$ is the variance of the distribution (other cases are discussed in the Supplemental Material~\cite{SupplementalMaterial}). With the addition of this small noise term, trajectories no longer approach the stable states asymptotically as in the deterministic case. Instead, a trajectory close to a stable state will  oscillate
stochastically within its basin of attraction, typically staying close to the fixed point for long periods of time. 
The trajectory will also make rare 
but large excursions from the stable state.
After sufficiently long time
an excursion
large enough to eject the trajectory from the original basin of attraction will necessarily occur, at which point it
 will transition to the neighborhood of another stable state of the system. The time scales
 for the occurrence of such transitions may be shorter or longer than the biologically relevant ones. The manipulation of these time scales underlies much of the control approach introduced below.

For a given noise intensity 
 $\varepsilon$, the transitions between two stable states, $i$ and  $j$, occur as a Poisson process with a certain rate $R^{\varepsilon} _{i,j}(\Omega)$. These rates can be computed by 
 evolving Eq.~\eqref{eqn:0a}, but in general at an unreasonably high computational cost. As a way to reduce this effort, we employ an asymptotic formula~\cite{Freidlin:2012vi}:
\begin{equation}
\label{eqn:1}
\tilde{R}^{\varepsilon}_{i,j} (\Omega)\propto \exp{ \left( - \frac{1}{\varepsilon} S^{*}_{i, j} (\Omega)\right) },
\end{equation}
where $\tilde{R}^{\varepsilon}_{i,j}$ serves as an excellent approximation to $R^{\varepsilon}_{i,j}$ for $\varepsilon$   small compared to $S^{*}_{i, j}$,
which is typically the case for noise associated with 
biophysical
systems. Here, $S^{*}_{i, j}$ is the minimum of the Freidlin-Wentzell action $S[ \,\cdot \, ]$, a functional over all possible transition paths $\vec\phi_{i,j}$ connecting the 
two stable states 
 in the state space. 
The path $\vec\phi^{*}_{i,j}$ minimizing the action for the given pair of stable states is 
calculated
numerically employing an implementation of the adaptive minimum action method~\cite{Zhou:2008jn},  in which 
we determine 
this minimum action path
through the optimization of a discretized version of the action functional using a quasi-Newton method  (Appendix~\ref{appendixA}).  
Conceptually, $S^{*}_{i, j}$ represents the cumulative height of all saddle points traversed by the minimum action path between 
the states $i$ and $j$.
It should be noted that a proportionality constant is omitted in the expression for $\tilde{R}^{\varepsilon}_{i,j}$. Although there is no known formula for computing this constant in general~\cite{Maier:1997vx},
the key condition for neglecting it is clear: 
as long as the actions $S_{i,j}$ associated with two paths differ by more than $\mathcal{O}(\varepsilon)$, the exponential term will dominate and the omission of the proportionality constant will not 
affect the result qualitatively.

The rate $\tilde{R}^{\varepsilon}_{i,j}$ represents the transition probability per unit of time along the most likely {\it direct} 
path(s) between the stable states $i$ and $j$. 
The transition rates through intermediate stable states can be determined by composing these elementary transition rates. It is often expected that the most likely transition path between two stable states would be a direct path, but as shown below, in many cases it  passes through intermediate stable states. Moreover, the transition paths and rates are generally asymmetric, with~$\tilde{R}^{\varepsilon}_{j,i}\neq \tilde{R}^{\varepsilon}_{i,j}$.
In the limit of long time, these transitions 
lead to an equilibrium probability distribution of occupied stable states,  
which we refer to as the {\it limiting occupancy} of the system and denote by $\vec{v}^{\,\varepsilon}$.  
Given that we generally study large populations of cells, in our applications it is convenient to 
interpret the limiting occupancy as an ensemble  average 
over different realizations of noise.

\begin{figure*}[t!]
\centerline{\includegraphics[width = 0.7\textwidth]{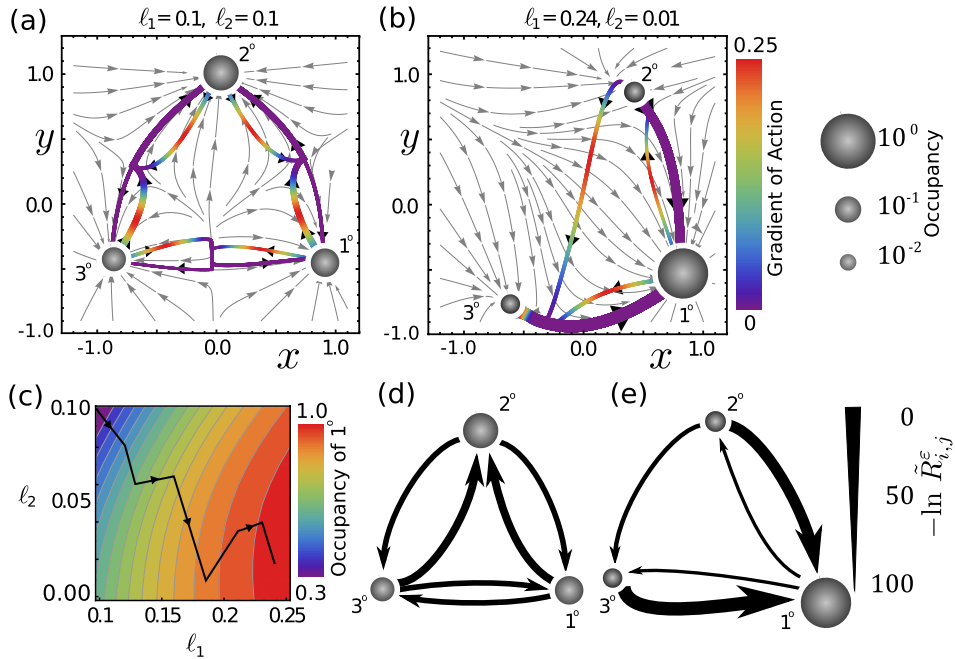}}
\caption{{ OLAC applied to a two-dimensional model of VPC differentiation.}
 (a, b) State space representation of stable states (nodes) and optimal transition paths (edges) before intervention (a) and after OLAC is applied to maximize the limiting occupancy of lineage $1^{\circ}$  (b).
 Node size indicates occupancy and edge width indicates the (negative of the) log transition rate along the corresponding 
 minimum action 
 path; the color code on  
 the transition paths indicates
 the derivative of the quasipotential. The background shows the velocity field for the given parameters.
For equal EGF and Notch signaling  (a), 
lineage $2^{\circ}$ is the one with highest occupancy.
The OLAC  solution
(b) indicates that high EGF signaling and low Notch signaling will lead to 
the maximum  
occupancy of lineage $1^{\circ}$. (c) Trajectory in the parameter space for one realization of the optimization procedure, where the contour plot indicates the limiting occupancy of lineage $1^{\circ}$. Note that each step of the optimization routine increases the occupancy of this state. (d, e) NESTs for the initial (d) and optimized (e) systems. 
In all panels, the noise intensity is  
assumed to be $\varepsilon = 0.007$ (as used previously~\cite{Corson:2012wm}).
}
\label{fig:1}
\end{figure*}

To illustrate the minimum action paths and their use in controlling cell behavior, we first consider a two-dimensional 
model for the \textit{Caenorhabditis elegans} vulval precursor cell (VPC) differentiation~\cite{Corson:2012wm}.
The VPC differentiation is representative of many other differentiation processes and enjoys significant experimental characterization. 
The model describes the differentiation of VPCs into one of three competent lineages, $1^{\circ}$, $2^{\circ}$, and $3^{\circ}$, corresponding to stable states in the system and marked 
as nodes 
 in Fig.~\ref{fig:1}(a). 
These lineages depend on two dimensionless parameters, $\ell_{1}$ and $\ell_{2}$,  that determine the levels of epidermal growth factor (EGF) and Notch signaling, respectively~(for the model equations, 
see Appendix~\ref{appendixC}).
It is known experimentally that increasing EGF and decreasing Notch signaling (relative to the base value of $0.1$) will bias cells towards lineage $1^{\circ}$, that decreasing EGF and increasing Notch will bias cells towards lineage $2^{\circ}$, and  that decreasing both EGF and Notch will bias cells towards lineage $3^{\circ}$~\cite{VD:2005dw}.
Figure~\ref{fig:1}(a) shows the state space for the VPC
system with equal, intermediate levels of EGF and Notch signaling ($\ell_{1} =\ell_{2}=0.1$) and a realistic level of noise ($\varepsilon = 0.007$). 
The calculated transition paths, transition rates, and limiting 
occupancies
are indicated by the edges, their width, and the size of the nodes, respectively.
For these parameters, our calculations indicate that the limiting occupancy is comparable for all three stable states, with lineage $2^{\circ}$ having the highest occupancy. 

\subsection{Optimal Least Action Control} 
We now turn to the manipulation of the tunable parameters $\Omega$, which
may include, for example,  gene expression levels, protein activity, and interaction rates.  Specifically, we seek to
modify $\Omega$ to optimize either specific transition rates $\tilde{R}^{\varepsilon}_{i,j}$ directly or the limiting occupancy $\vec{v}^{\,\varepsilon}$ through the alteration of the transition rates, 
while recognizing that the
parameters can be modified only within
specific ranges
determined by biophysical and experimental constraints.  
The latter include {\it sparsity} constraints, which we 
can use
to effectively limit the number of targets in the control set while avoiding a combinatorial explosion (Appendix~\ref{appendixD}). 
The problem is formalized as the  maximization of 
an objective function of interest $G$ (unique to each problem) over the parameters $\Omega$ subject to the 
given constraints $\{g_{k}(\Omega) =\zeta_{k}\}_{k}$ and  $\{h_{k}(\Omega) \le \eta_{k}\}_{k}$: 
\vspace{-0.1cm}
\begin{equation}
\max_{\substack{\Omega \\ \{g_{k}(\Omega) = \zeta_{k}\}_{k} \\ \{h_{k}(\Omega) \le \eta_{k}\}_{k}}} G\left(\{S^{*}_{i, j}(\Omega)\};\varepsilon\right). 
\label{equ3}
\vspace{-0.1cm}
\end{equation}
This procedure constitutes the central step of our implementation of OLAC. Note that this formulation is independent of the dimension and complexity of the system. Thus,
given a well-defined model we can identify control 
interventions
able to
alter the switching behavior and/or the  
stability of the states 
in desired ways 
without the need to know explicitly \textit{a priori} how variations of the tunable factors affect the system (beyond the implicit dependence defined by the dynamical equations). 

The objective function and the associated constraints do not need to be linear, allowing a wide range of possible dynamical behaviors to be optimized. 
 For any  
 set of such constraints,  Eq.~\eqref{equ3} can be solved numerically  through a second quasi-Newton method step that nests the adaptive minimum action method used to 
 determined the minimum action paths in connection with  Eq.~\eqref{eqn:1}.   Importantly, OLAC is highly scalable, with the following computational cost: 
\begin{equation}
Cost \sim \mathcal{O} \big(| \Omega | D (K+ \gamma P)\big), 
 \end{equation}
where $D$ is the number of variables defining the state space,  $| \Omega |$ is the number of tunable parameters under consideration, $P$ is the number of stable states, and $K$ is the number of transitions upon which $G$ depends.
This cost estimation follows from noting that every optimization step of OLAC relies on  $| \Omega |$ evaluations of $G$, each of which requiring $K$ runs of the nested optimization step, where each such run has a cost 
that is linear in $D$ when using the L-BFGS optimization method~\cite{Nocedal:2006uv}. 
Furthermore, for each top-level optimization step \eqref{equ3}, every one of the $P$ stable states has to be continued $| \Omega |$ times, each time at an integration cost that is 
linear in $D$ if the average network's degree is approximately constant, as in most network models (it would be at most quadratic in $D$ in the most general case)  (Appendix~\ref{appendixA}).   
Parameter 
$\gamma$ accounts for other constants describing the relative cost between optimization and integration.
Because of this high scalability, our method can be applied to complex high-dimensional multi-parameter systems without excessive computational cost. In this way OLAC  expands on previous foundational work that demonstrated how barriers between stable states can be altered to control the response to fluctuations and multistability~\cite{Dykman,Dykman6, Pisarchik, Rabitz, Schwartz1, Billings}
to now address large networks with many variables and many potential control parameters.

Before turning to high-dimensional systems,
we consider an illustrative example application of OLAC in which we maximize the final occupancy of lineage~$1^{\circ}$  in the VPC system. As an example constraint we stipulate that neither of the other two stable states be lost to bifurcation. This non-bifurcation constraint is intended to depict how the presence of dosing and/or experimental limitations may make complete elimination of undesired states infeasible.
This condition is imposed as the constraint that
the states $\vec{X}^*_k=\vec{X}^*_k(\Omega)$  representing each lineage $k$ remain stable fixed points:  
$\vec{g}_{k} (\Omega)\equiv \vec{F}(\vec{X}^*_k,\Omega) =\vec{0}$ 
and $h_k (\Omega) \equiv \mbox{Re}(\lambda_k) \le -\tau_0$.
In this notation,
$\lambda_k=\lambda_k(\Omega)$ is the eigenvalue of the Jacobian matrix
$D\vec{F}(\vec{X},\Omega)\big|_{\vec{X}^*_k}$
with largest real part, and  $\tau_0$ is a tolerance (set to be 0.1 in our simulations).
The objective function is $G(\Omega;\varepsilon)=v^{\varepsilon}_{1^{\circ}}(\Omega)$ and the tunable parameters are $\Omega=\{\ell_1,\ell_2\}$.
The result,  shown in Fig.~\ref{fig:1}(b), indicates a substantial (3-fold)  increase in the limiting occupancy of 
the lineage~$1^{\circ}$ state for  the optimal control intervention identified by OLAC. 
The parameter-space path for a representative realization of the optimization 
is shown in Fig.~\ref{fig:1}(c).
The optimal intervention, defined as an average over multiple realizations,  is a combination of increased EGF signaling ($\ell_{1}$: $0.1\to0.24$) and reduced Notch signaling ($\ell_{2}$:  $0.1 \to 0.01$), which has the net effect of lowering the barrier for transitions from lineages $2^{\circ}$ and $3^{\circ}$ to lineage $1^{\circ}$ while maintaining a high barrier for exiting this lineage. 
This controlled state corresponds to signaling strengths that have 
 been observed experimentally to indeed bias cell lineages towards lineage~$1^{\circ}$~\cite{VD:2005dw}. As mentioned above, for these results we utilized  Eq.~(\ref{eqn:1}) with the proportionality constants omitted to calculate transition rates between states. The inclusion of these prefactors could potentially alter the occupancy calculations of the stable states and in turn invalidate the results of OLAC. In the Supplemental Material~\cite{SupplementalMaterial}, Sec.~S1, we apply sensitivity analysis to quantify the uncertainty associated with omitting prefactors and show that doing so does not significantly alter the results in this case. That section also discusses more generally under which conditions prefactors can be omitted.
 \subsection{Network of State Transitions}   
 Our formulation
leads to a succinct and intuitive NEST representation for the transition dynamics, in which the nodes are unique stable states and the weighted, directed edges between nodes represent the rates of transition between the stable states. 
The basis for this representation is the observation that, for
small $\varepsilon$, the trajectories of the system in Eq.~\eqref{eqn:0a} are most often 
close to a stable state or transitioning between stable states along a minimum action path. The trajectories
will rarely venture 
into other regions of the state space,
 allowing us to focus only on this ``waiting-transition'' dynamics without sacrificing information about the system's behavior. 
For a system with $P$ stable states we can write the  $P\times P$ transition matrix  
\begin{equation}
\tilde{\textbf{R}}^{\varepsilon}_{i,j} = \begin{cases} \ \ \ \ \  \tilde{R}^{\varepsilon}_{i,j}  & i \neq j \\
 -\sum_{j} \tilde{R}^{\varepsilon}_{i,j}  & i = j \end{cases},
\end{equation}
which defines a (continuous-time) Markov process 
on
 the stable states of the
 system. 
 In our numerical calculations we use the fact that the limiting occupancy is described by the equilibrium solution of this process. 
 In fact, the Markov process specifies all attributes of the associated NEST.
The NEST is conceptually similar to the state transition network used in physical chemistry and 
 biochemistry~\cite{Becker97:dw,Rao2004,Reka,TransNetworks} as well as to transition networks studied in mathematics~\cite{Freidlin:2012vi}. 
There are, however, key differences between our approach and those considered previously, in addition to the fact that we focus on 
 network systems. In particular, the NEST does not assume the existence of a potential energy landscape and is defined for nonvanishing levels of noise, which required us to develop a new formulation that accounts in particular for long mixing times and for values of $\varepsilon$ larger than  $S^*_{i,j}$ (see Supplemental Material~\cite{SupplementalMaterial}, Sec.~S4). Furthermore, unlike static state transition networks, the transition rates in the NEST are malleable and can be rationally manipulated with OLAC. 

 \begin{figure*}
\centerline{\includegraphics[width = .7\textwidth]{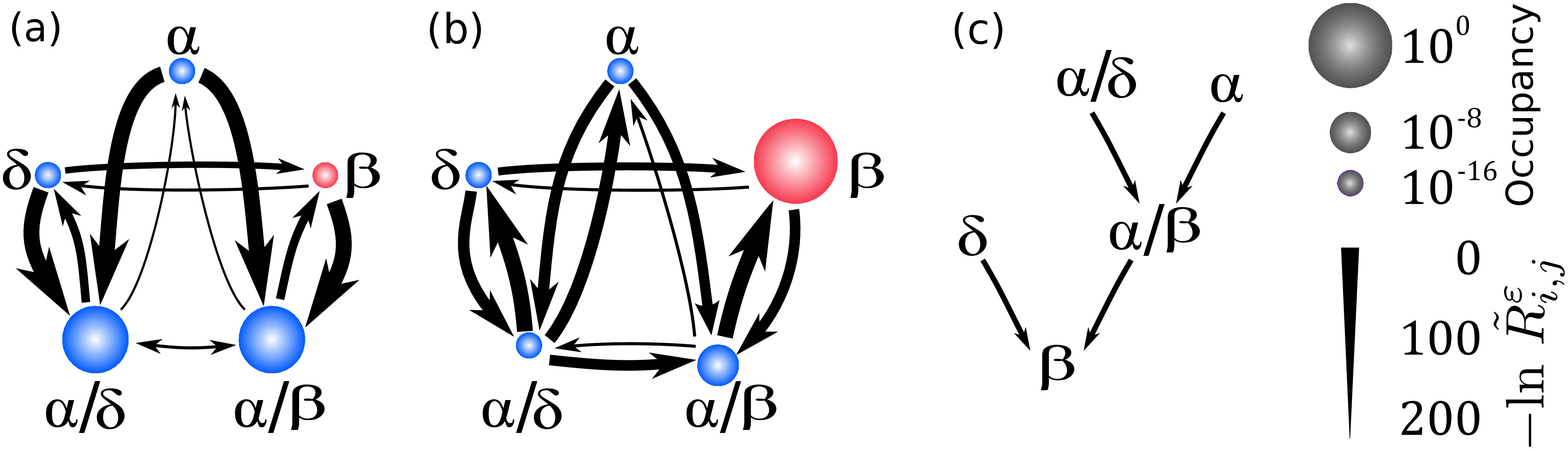}}
\caption{{ Controlled stochastic lineage switching in a multi-dimensional 
 model of HPC differentiation.} (a, b) NESTs of the model for the unmodified parameters  (a) and for OLAC applied to optimize the limiting occupancy of the $\beta$ cell state (red) (b), for $\varepsilon = 0.01$
 In both panels, node size indicates the limiting occupancy of the state and edge width indicates the (negative of the) log transition rate. OLAC identifies that only three (out of ten) transcription factors need to be tuned in the model to optimize $\beta$-cell  occupancy:  MafA (increased), Brn4 (decreased), and $\delta$-gene (decreased).
(c) Transition hierarchy into the $\beta$ cell state for the optimized system. Two states ($\delta$ and $\alpha/\beta$) transition directly; two others ($\alpha$ and $\alpha/\delta$) require passage through an intermediate state---in this case $\alpha/\beta$.  
}
\label{fig:3}
\end{figure*}

For the VPC model, the NESTs corresponding to the unmodified signaling strengths [Fig.~\ref{fig:1}(a)] and 
to the signaling strengths that optimize
lineage $1^{\circ}$ occupancy  [Fig.~\ref{fig:1}(b)] are shown in Figs.~\ref{fig:1}(d) and~\ref{fig:1}(e), respectively.  
These networks represent a substantial distillation of the dynamics of the underlying biophysical
system and can be used to simplify and explain the transition dynamics in a high-dimensional system without the need to consider its entire state space.
In particular,
for the range of edge widths shown,
the optimized NEST 
in Fig.~\ref{fig:1}(e) has 
edges between all nodes except from lineage $3^{\circ}$ to lineage $2^{\circ}$,  
 indicating that a direct transition between these two states is highly  
 unlikely
whereas an indirect transition is possible;
 indeed, the 
 two-step transition  
 $3^{\circ} {\to} 1^{\circ} {\to} 2^{\circ}$ 
 has  an overwhelmingly higher rate
 ($10^{2}$ times higher). 
 By comparing with the NEST of the original system, shown in Fig.~\ref{fig:1}(d), 
this also
 demonstrates
 that direct transitions for one parameter regime can become indirect for another.

The OLAC method can be 
implemented directly on the NEST representation.
Indeed, the objective function $G(\{S^{*}_{i, j}\};\varepsilon)$
is naturally defined in terms of the transition matrix of the system, $\tilde{\textbf{R}}^{\varepsilon}$. 
As our application to the VPC system shows, the combined effect of optimizing this objective function is to vary the height of the transition barrier along the minimum action path between the stable states.
This shows 
that OLAC itself does not require determining the full quasipotential of the system, which would be
computationally prohibitive in high dimensions.

\section{Applications}

\subsection{Controlling    
Pancreas Cell Transdifferentiation} 
An important  research problem
in cellular reprogramming concerns
 the induction of insulin producing pancreatic $\beta$ cells from non-insulin producing cell lineages; 
 interventions capable of
 achieving this goal could lead to new treatments for type I diabetes. In order to computationally identify  an
 optimal intervention to induce the desired reprogramming, 
 we consider a ten-dimensional model of the hierarchical pancreas cell (HPC) differentiation~\cite{Zhou:2011ju}.
 The model has
 five stable fixed points---three representing differentiated endocrine pancreas cell types ($\alpha$, $\beta$, and $\delta$) and two representing intermediate states ($\alpha$/$\beta$ and $\alpha$/$\delta$). The
 expression level of the ten 
 regulatory
genes, $\{ x_i \}$, 
 are each assumed to be tunable independently
 through a factor $\sigma_{i}$ (details of the model are given in Appendix~\ref{appendixC}).  
 
We first consider the uncontrolled 
 model, 
 in which $\sigma_{i} =1$ for $i =1, \dots, 10$. As shown in Fig.~\ref{fig:3}(a), in this case the two intermediate states attract the majority ($>99\%$) of  
 the occupancy.
  Furthermore, transitions to the $\beta$ state from the $\alpha$ and $\delta$ states occur at negligibly 
  small rates $(<10^{-100})$, 
  indicating that such lineage respecification 
  effectively 
 never occurs spontaneously. 
We apply OLAC to this model in order to identify the optimal combination of control actions that 
maximize the occupancy of the $\beta$ state, $v^{\varepsilon}_{\!\beta}$; 
 the admissible interventions are limited to  those  for which no bifurcations occur, which is imposed using  the same constraints as in the previous example.
The optimal intervention  
that maximizes
 $v^{\varepsilon}_{\beta}$ is a three-gene one, consisting of the downregulation of Brn4 and $\delta$-gene combined
with the upregulation of MafA.  
The resulting
NEST for $\varepsilon = 0.01$ 
is shown in Fig.~\ref{fig:3}(b). Under this optimal intervention, the limiting occupancy of the $\beta$ lineage 
goes from less than $ 0.01$ to  more than $0.99$.

The analysis specifically shows the reliance of  
some, but not all, lineages on
two-step
transitions to 
reach the desired $\beta$ lineage 
[Fig.~\ref{fig:3}(c)]. 
Previous research has suggested that indirect lineage respecifications 
might be suboptimal reprogramming strategies~\cite{Passier:2010dw}.
This is clearly the case for the $\delta\!\to\!\beta$ reprogramming, which is  optimized through a direct transdifferentiation event.
 For the $\alpha$ and $\alpha/\delta$ lineages, however, with overwhelming likelihood the transformation to the $\beta$ state will  pass through the intermediate $\alpha/\beta$ state.
Thus, which of the two cellular reprogramming strategies (direct or indirect) is optimal is context-dependent and cannot be determined without specification of the system, the initial state, and the final state.
Systematically accounting for such context-dependence could lead to new advances in the 
 development of 
 cellular reprogramming technologies. 
 
 \begin{figure*}[t!]
 \centerline{\includegraphics[width = 0.6\textwidth]{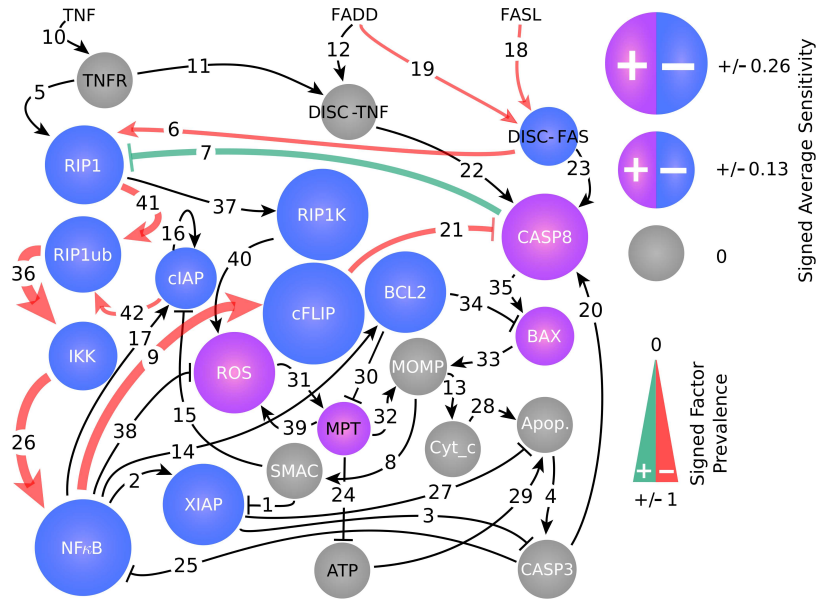}}                                                                                  
\caption{{ Optimal therapeutic interventions for the cell death regulatory network model.} The network has 22 genes (circles),  3 input parameters (top), and 42 target tunable parameters (edges), where the edge heads distinguish between excitatory (arrow) and inhibitory (bar) regulatory interactions. OLAC is applied to induce a bifurcation transition from 
survival to apoptotic states. The parameters involved in one or more optimal intervention are consistently upregulated (green) or downregulated (red) by the interventions; those not recruited by any optimal intervention are shown in black.  The edge width  
indicates
the prevalence of that parameter in  optimal  interventions for the various cell types and intervention strengths considered (up to the elimination of the survival state).  The size of each circle represents the sensitivity of the gene to pro-apoptotic interventions, defined as the change in gene expression between the  
uncontrolled
scenario and the smallest-intervention scenario at which the 
survival
state is eliminated; the color distinguishes up- and 
 down-expression.
Genes in white are those that change expression substantially between the 
survival
and apoptotic states. These results are based on 6 simulated cell types 
and 9 different  intervention strengths (Supplemental Material~\cite{SupplementalMaterial}, Fig.~S3).
}
\label{fig:4}
\end{figure*}

\subsection{Predicting Anti-Cancer Therapeutic Targets} 
Evasion of apoptosis is one of the hallmarks of cancer cells~\cite{hallmarks:2000dw}. 
As such, identifying tunable factors in the cell death pathway that 
effectively
eliminate proliferative (or abnormal survival) cell states without harming healthy cells could lead to 
new therapeutic targets. To computationally identify 
candidate targets, we employ OLAC to analyze a reformulation of the Boolean model of the cell death pathway proposed in Ref.~\cite{Calzone:2010dw}. This reformulation is a continuous-variable model generated using the  \emph{HillCube} 
methodology~\cite{Wittman:2009dw},  which is more amenable to analysis and  preserves all relevant properties of the original model, including the stable states. The model is comprised of $22$  genes central to the programmed cell death and 
$42$ parameters representing kinetic constants of the different reactions, which we denote by $c_i$, $i=1,\dots 42$  (Appendix~\ref{appendixC}). 
It follows that
two stable states form
 for all values of the parameters: {\it apoptosis} and {\it necrosis}. For a specific range of parameters representing healthy cells, a 
third state is also stable: the so-called  {\it naive} state. 
For a different parameter range, however, a different stable state arises,  
 corresponding to
 a  
 {\it survival}
  cell type that is resistant to the apoptotic signal.
This survival state
represents cancer and is the focus of our discussion.

Our goal is
to predict therapeutic targets that can 
induce transition from the survival state to the apoptotic state, without increasing the rate of
apoptotic death in normal cell types or causing them to become survival cells. 
Although noise can in principle induce switches from the survival to the apoptotic state, only a  fraction of the cells would transform as desired when both stable states exist. We therefore neglect the effect of noise temporarily and show that even in this case OLAC can be used to identify \emph{successful} optimal interventions, which under these conditions lead to a bifurcation that completely eliminates the  undesired (survival)
state.
To avoid inflammatory response, it is also desirable not to induce transitions to the necrotic state.
These conditions are assured by taking
 $G(\Omega)=-S^*$(survival $\rightarrow$ apoptosis) as the objective function to be maximized,  
and by imposing the
constraints $\Delta S^*$(naive $\rightarrow$ apoptotic) $\ge - \vartheta_0$, $\Delta S^*$(naive $\rightarrow$ necrotic) $\ge - \vartheta_0$, and  $S^*$(survival $\rightarrow$ necrotic) $\ge \vartheta_0$, where $\Delta$ indicates  change under the control intervention and $\vartheta_0$ is taken to be $0.05$ in our simulations.
 To encompass the largest possible set of candidate targets, we assume that any of the 
$42$ parameters of the model can be 
tuned in experiments, and hence we take $\Omega=\{c_i\}_{i=1}^{42}$.
However, since existing experimental techniques
cannot
be easily used to manipulate a large number of targets, we further constrain each control intervention to only involve a relatively small number of all tunable parameters. 
This can be
achieved by stipulating that 
the sum of the absolute values of all changes to the parameters must be equal to a pre-specified \emph{intervention strength}  $\chi_0$, as detailed in Appendix~\ref{appendixD}.

We thus apply OLAC to the cell death model for the objective function and constraints above.
To account for genetic heterogeneity between cancer cells, we analyze six different 
survival cell types,
represented here as six sets of unique 
values for the parameters $\Omega$. These parameters represent nondimensional kinetic constants for each gene-gene interaction. 
In our simulations the  individual
uncontrolled parameters for  these cell types  are on average  $0.50\pm0.03$ and
the intervention strength  $\chi_0$  is taken to be $0.1$, $0.2$, $\dots$ $0.9$ (Fig.~S3, in the Supplemental Material~\cite{SupplementalMaterial}, shows the breakdown of all cases).
The number of parameters modified by optimal interventions tends to increase as $\chi_0$ increases, 
ranging from an average of $2.7$ to $5.7$ for $\chi_0$ varied from $0.1$ to $0.9$. 
For each cell type, a successful intervention (eliminating the survival state) 
is always achieved for large enough $\chi_0$ 
within this interval. 
On  average, approximately only $5$ out of the $42$ 
parameters need to be modified in the optimal successful  interventions. To put this result in perspective, we note that if the sparsity constraint in Eq.~\eqref{eqnew} is disabled, OLAC leads to an average 
of no 
less than $40$ modified parameters for a  successful intervention.
This constraint, which generalizes immediately to any system, is therefore effective to restrain the number of control parameters.

Figure~\ref{fig:4} summarizes the results,
showing that 
a unique subset of only $10$ parameters is
needed to form the (on average) $5$-parameter successful target sets for any
cell type. The biological 
functions and prevalence of these targets are explicitly indicated in Table S1 of the Supplemental Material~\cite{SupplementalMaterial}. 
Notably, 
only two targets are included in interventions found for all six cell types, namely the parameters whose predicted 
increase will decrease the activation of NF$\kappa$B by IKK and the activation of IKK by RIP1ub (Fig.~\ref{fig:4}; parameters 26 and 36, respectively).  The identification of these two targets  is 
not entirely surprising since NF$\kappa$B is a central regulator of the cell death pathway whose over-activation has been implicated in the cellular transition to cancer~\cite{Luo:2005dw};
the consistent
identification of both targets across all six 
cell types 
is an indication of
 the robustness of our approach.
Aside from these global 
targets, OLAC also predicts unique combinations of targets for each cell type, in many cases indicating genes and interactions that have only recently been identified as possible cancer targets---e.g., the potential of suppressing the activation of cFLIP by NF$\kappa$B~\cite{Safa:dw2012} (Fig.~\ref{fig:4}; parameter 9). The identification of  optimal 
target combinations that are unique to different cell types illustrates the potential of OLAC to assist the development of personalized therapeutic strategies 
as well as of interventions to address various forms of cancer~\cite{Saadatpour2011} and to manipulate heterogeneous multistable cells in general~\cite{Regan2012,Lang,Reka3}.

OLAC finds an optimal control action whether or not a bifurcation has been reached, allowing its efficacy and possible adverse effects to be monitored in experiments as the strength of the intervention is increased. Theoretically, the efficacy of an intervention can be defined as the relative reduction of the action associated with the transition from the survival state to the apoptotic state [Supplemental Material~\cite{SupplementalMaterial}, Fig.~S3(c)]. Experimentally, the efficacy can be more easily estimated by monitoring the 
predicted gene expression changes induced by the interventions [Supplemental Material~\cite{SupplementalMaterial}, Fig.~S3(b)].
As shown in Fig.~\ref{fig:4} for interventions that eliminate the survival state, the sensitivity to the control interventions can vary widely across different genes.
For example, in every cell type considered, the expressions of CASP3, SMAC, and CYT-c (among others) are predicted not to change at all until the elimination of the survival state; 
in contrast, cFLIP, IKK, and NF$\kappa$B  
change expression in  all cell types. 

\begin{figure}      
\centerline{\includegraphics[width = .45\textwidth]{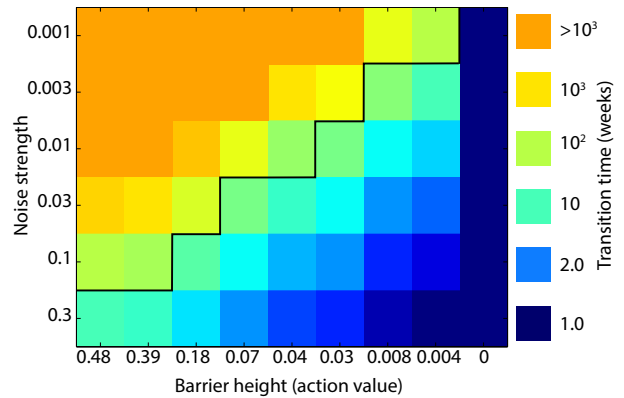}}                                                                                  
\caption{Transition time for the cell death model as a function of barrier height and noise strength. Transition time was calculated using Eq.~(\ref{eqn:Exittime}).  
The black line separates those barrier height/noise strength combinations that occur over a therapeutically relevant time scale (bottom right) and those that do not (top left).}
\label{fig:5}
\end{figure}

\subsection{Biophysically Relevant Transition Times}  
 In cases where OLAC identifies an intervention to induce a bifurcation to eliminate 
an undesired state, it is expected from experimental work in deterministically reprogramming somatic cells to a pluripotent state (a prototypical example of an induced cell state transition) that such a transition will occur on a relatively short time scale, within approximately one week~\cite{Hanna2}. However, due to constraints on parameter changes, it may not always be possible to  induce a bifurcation.
To maintain biological 
relevance, the time scale over which these non-bifurcative interventions act should not be exceedingly long. Given a noise strength $\varepsilon$ and a barrier height $S^{*}_{i,j}$ between two states, the approximate mean first exit time $\tilde{\mathcal{T}}^{\varepsilon}_{i,j}(\Omega)$ from that state can be estimated as~\cite{Freidlin:2012vi} 
\begin{equation}
\label{eqn:Exittime}
\tilde{\mathcal{T}}^{\varepsilon}_{i,j}(\Omega) \propto \exp \left( \frac{1}{\varepsilon} S^{*}_{i,j}(\Omega) \right),
\end{equation} 
where the transition time increases for lower noise levels and higher action barriers, as expected.
Since $\tilde{\mathcal{T}}^{\varepsilon}_{i,j}$ in Eq.~(\ref{eqn:Exittime}) is dimensionless, this quantity is best interpreted as a relative increase in transition time over the case of $S^{*}_{i,j}=0$, which from above we take to be one week. 

Figure~\ref{fig:5} shows the mean first transition time using as a model application the cell death model considered in the previous section.  The figure shows the mean transition time as a function of both noise level and barrier heights  for a single cell  type (as defined by each intervention strength in Fig.~S3(c) of the Supplemental Material~\cite{SupplementalMaterial}).
It follows that 
for cases where $S^{*}_{i,j}(\Omega) \leq 3 \varepsilon$, the average transition time will be less than 20 weeks. This time scale is a reasonable upper limit for the
 biological 
relevance of any intervention:  because transition times are exponentially distributed, interventions at 
this strength will cause a measurable fraction of the population to transition in just a couple of weeks.
This benchmark thus provides a straightforward criterion to determine if an identified intervention will be relevant in practice. 
Figure~\ref{fig:5} also demonstrates the 
important
fact that it is not the size of the dynamical system that determines the switching time between states, but rather the 
ratio of the barrier height 
to the strength of the noise.

We expand on this approach in the Supplemental Material~\cite{SupplementalMaterial}, Sec.~S3, where we demonstrate that OLAC can be used to identify 
constrained 
interventions that achieve a desired limiting occupancy within a prespecified time frame. In that section we also generalize OLAC to systems with (multiplicative) noise that depends on the system state and to 
systems that are best modeled using a chemical kinetic formulation~\cite{StochRev, DiLiu1, DiLiu2}.

\section{Discussion}
The method proposed here,  optimal least action control (OLAC), represents a new 
direction
in the control of  biophysical 
systems.
Traditional control approaches for biophysical systems are based on manipulating the trajectories of the system, while our method is based instead on manipulating
the epigenetic landscape through which  these trajectories travel. This is achieved by 
effectively
altering the barriers between different stable states in the quasipotential of the system, which could be achieved biologically through, for example, a genome editing 
approach~\cite{GenomeEditing}.
In this way OLAC  stands in contrast to other orthogonal methods that 
seek to control cellular states by directly modifying gene expression levels through, for example,  siRNA strategies~\cite{Sean, RealsiRNA, AlbertArx}. As such, our method naturally accounts for cellular noise and the incorporation of constraints on the possible control actions.
But in contrast to previous work that has sought to construct the entire quasipotential for a dynamical 
system~\cite{Wang:2011ef},
 we  
 utilize
 the associated {\it network of state transitions} (NEST)
 to describe and control the system at a substantially reduced computational cost, which renders the approach applicable to a wide range of biophysical systems---including high-dimensional networks.

The cell death model
example illustrates one of the key strengths of OLAC: its  
effectiveness when the problem involves an 
 explosion in the number of possible reprogramming combinations. 
 In practical applications to multi-parameter systems, it is of interest to identify interventions that are not only optimal but also sparse---i.e.,  that target only a few out of many possible tunable factors. Such sparsity is desired since most biologically realizable interventions are able to control only a few parameters and, for example, would not be able to directly control all  
 parameters in the cell death model. 
Because OLAC 
can benefit from the framework of convex regularization,
however, this combinatorial increase in the number of possible interventions can be dealt with easily by incorporating a sparsity constraint that has only marginal impact on the computational cost.

Another key property of the approach is its flexibility.
For example, previous modeling approaches to explain reprogramming experiments have focused mainly on bifurcations  that destabilize or eliminate states~\cite{Chickarmane:2008ha},  which are comparatively larger changes in the dynamics of the system.
These approaches  do not benefit from the presence of noise and 
tacitly assume that, to reach the desired state, the initial state must become deterministically unstable or disappear, which,
 as demonstrated for the cell death model, may not be possible given a stringent enough set of constraints on the possible set of interventions.
On the other hand, for being based on the  Freidlin-Wentzell action, our method 
is effective as a unified approach both (i) to exploit the presence of noise for control in the absence of any bifurcation (as shown for the pancreas model and  the cell death model with low intervention strengths) and (ii) to identify control interventions mediated by bifurcations in the absence of any noise (as shown for the cell death model with large intervention strengths). 
Therefore, instead of facing reduced performance in the presence of noise, which is a common drawback of other control approaches~\cite{Sean}, OLAC benefits from the presence of noise, utilizing noise as an additional control tool.

Through the use of 
the approach introduced here
we have shown that, counterintuitively,  
the optimal lineage respecification trajectory is often indirect; that is, 
they correspond to cases in which 
the most likely trajectory for an optimized transition between two states will
 pass through one or more intermediate stable states. 
Such cases cannot be anticipated by common sense since for a therapeutic intervention, for instance,  
an indirect path may cause the cells to worsen before they improve. 
This result suggests a new possible method for identifying enhanced reprogramming strategies, namely by systematically 
exploring
combinations of 
intermediate transitions. 

Our approach can also be applied to a much broader range of biophysical 
problems than those discussed in detail here. In particular, 
OLAC could be used in the
context of synthetic biology, as researchers seek to build ever more complex
synthetic systems and
computer-aided design methods  
play an increasingly important role~\cite{Clancy:2010jv}. 
In that context, OLAC can identify optimal parameter tuning 
to reshape the quasipotential 
for the rational design 
of systems with 
pre-specified dynamical behavior and response to noise.
The same approach can also be used to generate insights into epigenetic diseases and the mechanisms that give rise to them, including the possible dependence of their incidence rate on external versus genetic factors, as well as insights into potential preventive measures to reduce disease risks by identifying conditions that increase the barriers for transitions to disease states.

Furthermore, the foundation of OLAC in the Freidlin-Wentzell action means that the applications of this method need not be biophysical. In particular, our method can easily be used to predict control interventions in other noisy multistable networks and, along with NEST, to characterize the basins of attraction of these systems~\cite{Kurths}. For example, in the Supplemental Material~\cite{SupplementalMaterial}, Sec.~S4, we construct the NEST for a dynamical system with more than 100 attractors~\cite{Feudel2008} and demonstrate that even in a system with such substantial multistability, noise can effectively eliminate the occupancy of the majority of attractors, leaving only a small fraction of them occupied. Moreover, along with the already mentioned applications to power-grids, polymer networks, and food-web networks, OLAC could also find use in controlling spreading processes on social networks~\cite{Granovetter}, 
in inducing
synchronization patterns 
in oscillator networks~\cite{Abrams}, in manipulating associative memory networks~\cite{Hopfield1}, and 
potentially in creating new attractors~\cite{Campbell}. Further development of this method could also expand its applications to models of disease epidemics and population dynamics. In particular, substantial foundational work has been done on the modeling of extinction events in such systems, which typically requires model-specific mathematical analysis \cite{Assaf1,DykmanSchwartz1,Dykman1}. Because a white-noise approximation, like in Eq.~(\ref{eqn:0a}), cannot accurately approximate the dynamics of these systems \cite{Doering1}, expanding OLAC to apply to extinction events in larger networks will require using situation-specific calculations of the transition rates.

Ultimately, we believe that OLAC---together with 
 NEST---forms a
flexible, scalable method which can be used to understand and control the dynamical 
 stability and response to noise of 
 a wide range of 
 complex networks, including those with large number of dynamical variables, tunable parameters, and attractors. The method is easily implementable,  with  a ready-to-use numerical implementation included as supplemental files \cite{SupplementalMaterial}.
 The method 
 requires no \textit{a priori} information (beyond those implicitly defined by the dynamical equations) about how variations in the control
parameters affect the system, and it can be used in concert with 
rather general
constraints on the control actions.  While  OLAC can be applied to  many systems, as formulated here application of the method requires a quantitative dynamical model. Extending the method to systems for which no model is available is an important direction for future research.
Future  work is also expected to expand on the applications of the approach and to 
further demonstrate its  experimental efficacy.

\begin{acknowledgments}
The authors thank Jorge Nocedal for insightful discussions on optimization methods, and  members of the Leonard and Thomas-Elliott labs for continued support.
This work was funded by the National Cancer Institute Grants No. 1U54CA143869 (NU PS-OC) and No. 1U54CA193419 (Chicago area PS-OC),  the National Institute of General Medical Sciences Grant R01GM113238, the National Science Foundation  Grant No. DMS-1057128, the Chicago Biomedical Consortium, and a National Science Foundation Graduate Research Fellowship. 
\end{acknowledgments}

\appendix

\section{Calculating the Minimum Action Value} 
\label{appendixA}

The minimum action $S^{*}_{i, j}$ is determined using 
 \begin{equation}
\begin{aligned}
\label{eqn:Actions}
S^{*}_{i, j} &= S[\vec{\phi}^{*}_{i,j}]= \min_{\substack{\vec{\phi}(t) \\ \vec{\phi}(T_{1})=\vec{a}_{i}  \\ \vec{\phi}(T_{2})=\vec{a}_{j}}} \left(S[\vec{\phi}] \right), \\
 \quad S[\vec{\phi}] &= \frac{1}{2}\int_{T_{1}}^{T_{2}} \left \| \frac{d \vec{\phi}}{dt}(t) - \vec{F}(\vec{\phi}(t); \Omega) \right \|^{2} dt,
\end{aligned}
\end{equation}
where $\vec{a}_{i}$ and $\vec{a}_{j}$ denote the coordinates of the initial and final stable states, respectively. 
 To solve this optimization problem we minimize the discretized version of the functional~\cite{Zhou:2008jn}, given by
\begin{multline}
 S_{t_{0}, \dots, t_{m_s}} \left[ \vec{\Phi}_{0}, \dots, \vec{\Phi}_{m_s} \right] = \\\frac{1}{2} \sum_{n=1}^{m_s} \left \| \frac{\vec{\Phi}_{n} - \vec{\Phi}_{n-1}}{\Delta t_{n}} - \vec{F}(\vec{\Phi}_{n - 1/2}; \Omega) \right \|^{2} \Delta t_{n},
 \label{eqn:3s}
\end{multline}
where we use $T_{1} = t_{0} < t_{1} < \dots < t_{m_s} = T_{2}$, $\Delta t_n=t_n-t_{n-1}$,
$\vec{\Phi}_{n} = \vec{\phi}(t_{n})$, and $\vec{\Phi}_{n-1/2} = (\vec{\Phi}_n + \vec{\Phi}_{n-1})/2$. 
In our simulations
we set $T_{2} = -T_{1} = 20$ and $m_s=100$, 
and verified that larger values of $T_{2}  -T_{1}$ and $m_s$  would not improve accuracy noticeably. 

To maximize efficiency, we regularly remesh the path from the time domain to the space domain and adaptively redefine $t_{n}$ according to 
\begin{equation}
\Delta \alpha_{n}
\int_{T_{1}}^{T_{2}}w(t) dt = w(t_{n-1/2})\Delta t_{n},
\end{equation}
where $\Delta \alpha_{n} = 1/m_s$, 
$t_{n-1/2} = (t_n + t_{n-1})/2$,  and $w(t) = \sqrt{1+ C \| \frac{d\vec{\phi}}{dt}(t) \|^{2}}$ is the monitor function measuring the speed along the path. We used the initial $t_{n}$ evenly spaced and $C=10^{10}$.
These calculations require an initial path between the stable states. 
The numerical results we report  are obtained using a straight line as the initial path.
We have checked 
that using different initial paths, such as those generated through the Brownian bridge approach~\cite{Karatzas:1991ws}, the simulation 
always converges to the same optimal paths. 

We note that, since the parameters of the  
system 
are modified at each step of OLAC, robust numerical continuation of the equilibria is necessary.  
We use a simple homotopy method to continue the stable states from  initial parameters $\Omega'$ to terminal  parameters $\Omega''$.
Specifically, we use linear interpolation between  
these parameter values
combined with a Newton step~\cite{Seydel:1988vz}.
The latter includes
checking at each step that the desired states are  not lost to unintended bifurcations.

 All optimization procedures are done employing the interior-point algorithm in the  \verb|fmincon| function of the optimization toolbox of MATLAB~\cite{2dw}.
A MATLAB implementation of OLAC in included as supplemental files (see Supplemental Material \cite{SupplementalMaterial}, Sec.~S2, for details).

\section{Equations of the Models Considered}
\label{appendixC}

\textit{VPC Differentiation Model.}\ The deterministic component of the VPC model~\cite{Corson:2012wm} is given by
\begin{equation}
\frac{d \vec{r}}{dt} = \frac{1}{\tau} ( \vec{\sigma}_{1} - \vec{r}) , 
\end{equation}
where $\vec{r} = (x, y)^{T}$, $\vec{\sigma}_{1} = \tanh \big( \| \vec{f} + \vec{m} \| \big) \frac{\vec{f} + \vec{m}}{\| \vec{f} + \vec{m} \|}$,  $\vec{f} = \left[2x - 2c_{2} xy, 2y + c_{2} (y^{2}-x^{2} ) \right]^{T}$, and $\vec{m} = \vec{m}_{0} + \ell_{1} \vec{m}_{1} +\ell_{2}\vec{m}_{2}$. This is an effective representation of the combined effects of EGF and Notch signaling, whose signaling strengths are determined by $\ell_{1}$ and $\ell_{2}$, respectively. The other parameters are $\tau = 2.18$, $\vec{m}_{0} = (-0.86, -0.50)^{T}$, $\vec{m}_{1} = (0.86, -0.50)^{T}$, and $\vec{m}_{2} = (0.0, 1.0)^{T}$.

\textit{HPC Differentiation Model.} The deterministic component of the HPC differentiation model is~\cite{Zhou:2011ju}
\begin{equation}
\begin{aligned}
\mbox{Pdx1:} \quad& \frac{d {x}_{1}}{dt} = \sigma_{1} a_{s}  \frac{x_{4}^{n} + x_{7}^{h} + x_{8}^{h} + x_{10}^{h}}{1 +  x_{4}^{h} + x_{7}^{h} + x_{8}^{h} + x_{10}^{h}} - k_{deg}x_{1},  &      \\
\mbox{Ptf1a:} \quad&\frac{d {x}_{2}}{dt} = \sigma_{2} a \frac{x_{10}^{h}}{1 + x_{10}^{h} + x_{3}^{h}} - k_{deg}x_{2}, &\\
\mbox{Ngn3:} \quad&\frac{d {x}_{3}}{dt} = \sigma_{3} a \frac{x_{10}^{h}}{1 + x_{10}^{h} + x_{2}^{h}} - k_{deg}x_{3},&\\
\mbox{Pax6:} \quad&\frac{d {x}_{4}}{dt} = \sigma_{4} a\frac{\mu^{h} x_{3}^{h} + x_{4}^{h}}{ 1 + \mu^{h}x_{3}^{h} + x_{4}^{h}} - k_{deg}x_{4},&\\
\mbox{Pax4:} \quad&\frac{d {x}_{5}}{dt} = \sigma_{5} a_{e} \frac{\mu^{h}_{m} x_{1}^{h} \mu^{h} x_{3}^{h}}{1 + \mu^{h}_{m} x_{1}^{h} \mu^{h} x_{3}^{4} + x_{6}^{h}} - k_{deg}x_{5},&\\
\mbox{Arx:} \quad&\frac{d {x}_{6}}{dt} = \sigma_{6} a_{e} \frac{\mu^{h}_{m} x_{1}^{h} \mu^{h} x_{3}^{h}}{1 +\mu^{h}_{m}x_{1}^{h} \mu^{h} x_{3}^{h} + x_{5}^{h}} - k_{deg}x_{6}, &\\
\mbox{MafA:} \quad&\frac{d {x}_{7}}{dt} = \sigma_{7}  a \frac{\mu^{h}_{m} x_{1}^{h} \mu^{h} x_{3}^{h} x_{5}^{h} + x_{7}^{h}}{1 + \mu^{h}_{m} x_{1}^{h} \mu^{h} x_{3}^{h} x_{5}^{h} + x_{7}^{h} + x_{8}^{h}} -k_{deg} x_{7},&\\
\delta \mbox{-gene:} \quad&\frac{d {x}_{8}}{dt} = \sigma_{8} a \frac{\mu^{h}_{m} x_{1}^{h} \mu^{h} x_{3}^{h} x_{5}^{h} + x_{8}^{h}}{1 + \mu^{h}_{m} x_{1}^{h} \mu^{h} x_{3}^{h} x_{5}^{h} + x_{7}^{h} + x_{8}^{h}} - k_{deg}x_{8},&\\
\mbox{Brn4:} \quad&\frac{d {x}_{9}}{dt} = \sigma_{9}  a \frac{\mu^{h} x_{6}^{h} + x_{9}^{h}}{1 + \mu^{h} x_{6}^{h} + x_{9}^{h}} - k_{deg}x_{9},\\
\mbox{Hnf6:} \quad&\frac{d {x}_{10}}{dt} =  - \sigma_{10} k_{deg}x_{10},&\\     
\end{aligned}
\end{equation}
where $(a_{s}, a_{e}, a, k_{deg}, \mu, \mu_{m}, h) = (2.2, 6.0, 4.0, 1.0, 0.25,$ $0.125,4.0)$ are the fixed parameters in these equations. Each parameter $\sigma_{i}$ represents a tunable factor to alter the expression of the gene represented by $x_{i}$.

\textit{Cell Death Model.} The cell death model was converted from the Boolean model~\cite{Calzone:2010dw} (available at 
\url{http://www.ebi.ac.uk/biomodels-main/ MODEL0912180000}) into a continuous version 
using the Odefy package~\cite{Wittman:2009dw}. 
The system has 22 variables, representing gene products, and 42 tunable parameters.
In addition, the model has 3 input parameters that  do not change in time, and 3 output variables that indicate 
the state of the cell (distinct combinations of which indicate whether the cell is in the survival, apoptotic,  necrotic, or naive state).
\section{Implementing the Sparsity Constraint} 
\label{appendixD}
In many biological systems there are dozens or hundreds of parameters that could potentially be changed, but in general only a few of them can be changed in any one intervention. To identify the few most promising targets from a large field of possible ones we employ a sparsity constraint. The constraint
is implemented  as  
\vspace{-0.2cm}
 \begin{equation}
 \sum_{i=1}^{|\Omega|} \big| \Delta \Omega_i \big| \leq \chi_0,
 \label{eqnew}
 \vspace{-0.2cm}
 \end{equation} 
where, as above, 
$ \Delta$ denotes the change due to the control action.
While  
the condition in Eq.~\eqref{eqnew}
is by itself consistent with all parameters being altered, optimization under this constraint (as the one invoked by OLAC) is expected to lead to a reduced number of modified parameters. 
The basis for this conclusion is that
this constraint works similarly to well-established methods of convex regularization, which are known to lead to sparsity under  general conditions~\cite{6dw}.
The specific number of modified parameters as well as success rate will generally depend on $\chi_0$, and this dependence can be explored as an additional control factor.
This formulation has the remarkable 
advantage of involving only one optimization step, and hence avoids the combinatorial explosion that would be involved in testing all 
$\frac{|\Omega|!}{k!(|\Omega|-k)!}$  combinations of possible ``$|\Omega|$ choose $k$'' tunable factors.
Indeed, an exhaustive 
strategy 
would be computationally prohibitive
since testing for all $k$ would require $2^{|\Omega|}-1$ optimization steps, which is $>10^{12}$ for $|\Omega|=42$.
Naturally,  if a particular parameter is 
not
targetable under the given conditions, such information can be directly 
be incorporated into our analysis.

\input{Manuscript.bbl}
\end{document}

%% file: Manuscript.bbl
\providecommand{\noopsort}[1]{}\providecommand{\singleletter}[1]{#1}%

%% file: Manuscript.bbl
\begin{thebibliography}{72}%
\makeatletter
\providecommand \@ifxundefined [1]{%
 \@ifx{#1\undefined}
}%
\providecommand \@ifnum [1]{%
 \ifnum #1\expandafter \@firstoftwo
 \else \expandafter \@secondoftwo
 \fi
}%
\providecommand \@ifx [1]{%
 \ifx #1\expandafter \@firstoftwo
 \else \expandafter \@secondoftwo
 \fi
}%
\providecommand \natexlab [1]{#1}%
\providecommand \enquote  [1]{\emph{#1}}%
\providecommand \bibnamefont  [1]{#1}%
\providecommand \bibfnamefont [1]{#1}%
\providecommand \citenamefont [1]{#1}%
\providecommand \href@noop [0]{\@secondoftwo}%
\providecommand \href [0]{\begingroup \@sanitize@url \@href}%
\providecommand \@href[1]{\@@startlink{#1}\@@href}%
\providecommand \@@href[1]{\endgroup#1\@@endlink}%
\providecommand \@sanitize@url [0]{\catcode `\\12\catcode `\$12\catcode
  `\&12\catcode `\#12\catcode `\^12\catcode `\_12\catcode `\%12\relax}%
\providecommand \@@startlink[1]{}%
\providecommand \@@endlink[0]{}%
\providecommand \url  [0]{\begingroup\@sanitize@url \@url }%
\providecommand \@url [1]{\endgroup\@href {#1}{\urlprefix }}%
\providecommand \urlprefix  [0]{URL }%
\providecommand \Eprint [0]{\href }%
\providecommand \doibase [0]{http://dx.doi.org/}%
\providecommand \selectlanguage [0]{\@gobble}%
\providecommand \bibinfo  [0]{\@secondoftwo}%
\providecommand \bibfield  [0]{\@secondoftwo}%
\providecommand \translation [1]{[#1]}%
\providecommand \BibitemOpen [0]{}%
\providecommand \bibitemStop [0]{}%
\providecommand \bibitemNoStop [0]{.\EOS\space}%
\providecommand \EOS [0]{\spacefactor3000\relax}%
\providecommand \BibitemShut  [1]{\csname bibitem#1\endcsname}%
\let\auto@bib@innerbib\@empty
\bibitem [{\citenamefont {Raj}\ and\ \citenamefont {van
  Oudenaarden}(2008)}]{Raj:2008ip}%
  \BibitemOpen
  \bibfield  {author} {\bibinfo {author} {\bibfnamefont {A.}~\bibnamefont
  {Raj}}\ and\ \bibinfo {author} {\bibfnamefont {A.}~\bibnamefont {van
  Oudenaarden}},\ }\bibfield  {title} {\enquote {\bibinfo {title} {{Nature,
  Nurture, or Chance: Stochastic Gene Expression and Its Consequences}},}\
  }\href@noop {} {\bibfield  {journal} {\bibinfo  {journal} {Cell}\ }\textbf
  {\bibinfo {volume} {135}},\ \bibinfo {pages} {216} (\bibinfo {year}
  {2008})}\BibitemShut {NoStop}%
\bibitem [{\citenamefont {McAdams}\ and\ \citenamefont
  {Arkin}(1997)}]{Arkin:1997dw}%
  \BibitemOpen
  \bibfield  {author} {\bibinfo {author} {\bibfnamefont {H.~H.}\ \bibnamefont
  {McAdams}}\ and\ \bibinfo {author} {\bibfnamefont {A.}~\bibnamefont
  {Arkin}},\ }\bibfield  {title} {\enquote {\bibinfo {title} {{Stochastic
  Mechanisms in Gene Expression}},}\ }\href@noop {} {\bibfield  {journal}
  {\bibinfo  {journal} {Proc. Natl. Acad. Sci. U.S.A.}\ }\textbf {\bibinfo
  {volume} {95}},\ \bibinfo {pages} {814} (\bibinfo {year} {1997})}\BibitemShut
  {NoStop}%
\bibitem [{\citenamefont {Balazsi}\ \emph {et~al.}(2011)\citenamefont
  {Balazsi}, \citenamefont {van Oudenaarden},\ and\ \citenamefont
  {Collins}}]{Balazsi:2011id}%
  \BibitemOpen
  \bibfield  {author} {\bibinfo {author} {\bibfnamefont {G.}~\bibnamefont
  {Balazsi}}, \bibinfo {author} {\bibfnamefont {A.}~\bibnamefont {van
  Oudenaarden}}, \ and\ \bibinfo {author} {\bibfnamefont {J.}~\bibnamefont
  {Collins}},\ }\bibfield  {title} {\enquote {\bibinfo {title} {{Cellular
  Decision Making and Biological noise: From Microbes to Mammals}},}\
  }\href@noop {} {\bibfield  {journal} {\bibinfo  {journal} {Cell}\ }\textbf
  {\bibinfo {volume} {144}},\ \bibinfo {pages} {910} (\bibinfo {year}
  {2011})}\BibitemShut {NoStop}%
\bibitem [{\citenamefont {Chang}\ \emph {et~al.}(2008)\citenamefont {Chang},
  \citenamefont {Hemberg}, \citenamefont {Barahona}, \citenamefont {Ingber},\
  and\ \citenamefont {Huang}}]{Chang:2008gu}%
  \BibitemOpen
  \bibfield  {author} {\bibinfo {author} {\bibfnamefont {H.~H.}\ \bibnamefont
  {Chang}}, \bibinfo {author} {\bibfnamefont {M.}~\bibnamefont {Hemberg}},
  \bibinfo {author} {\bibfnamefont {M.}~\bibnamefont {Barahona}}, \bibinfo
  {author} {\bibfnamefont {D.~E.}\ \bibnamefont {Ingber}}, \ and\ \bibinfo
  {author} {\bibfnamefont {S.}~\bibnamefont {Huang}},\ }\bibfield  {title}
  {\enquote {\bibinfo {title} {{Transcriptome-Wide Noise Controls Lineage
  Choice in Mammalian Progenitor Cells}},}\ }\href@noop {} {\bibfield
  {journal} {\bibinfo  {journal} {Nature (London)}\ }\textbf {\bibinfo {volume}
  {453}},\ \bibinfo {pages} {544} (\bibinfo {year} {2008})}\BibitemShut
  {NoStop}%
\bibitem [{\citenamefont {Choi}\ \emph {et~al.}(2008)\citenamefont {Choi},
  \citenamefont {Cai}, \citenamefont {Frieda},\ and\ \citenamefont
  {Xie}}]{Choi:2008ke}%
  \BibitemOpen
  \bibfield  {author} {\bibinfo {author} {\bibfnamefont {P.~J.}\ \bibnamefont
  {Choi}}, \bibinfo {author} {\bibfnamefont {L.}~\bibnamefont {Cai}}, \bibinfo
  {author} {\bibfnamefont {K.}~\bibnamefont {Frieda}}, \ and\ \bibinfo {author}
  {\bibfnamefont {X.S.}\ \bibnamefont {Xie}},\ }\bibfield  {title} {\enquote
  {\bibinfo {title} {{A Stochastic Single-Molecule Event Triggers Phenotype
  Switching of a Bacterial Cell}},}\ }\href@noop {} {\bibfield  {journal}
  {\bibinfo  {journal} {Science}\ }\textbf {\bibinfo {volume} {322}},\ \bibinfo
  {pages} {442} (\bibinfo {year} {2008})}\BibitemShut {NoStop}%
\bibitem [{\citenamefont {Gupta}\ \emph {et~al.}(2011)\citenamefont {Gupta},
  \citenamefont {Fillmore}, \citenamefont {Jiang}, \citenamefont {Shapira},
  \citenamefont {Tao}, \citenamefont {Kuperwasser},\ and\ \citenamefont
  {Lander}}]{Gupta:2011fk}%
  \BibitemOpen
  \bibfield  {author} {\bibinfo {author} {\bibfnamefont {P.~B.}\ \bibnamefont
  {Gupta}}, \bibinfo {author} {\bibfnamefont {C.~M.}\ \bibnamefont {Fillmore}},
  \bibinfo {author} {\bibfnamefont {G.}~\bibnamefont {Jiang}}, \bibinfo
  {author} {\bibfnamefont {S.~D.}\ \bibnamefont {Shapira}}, \bibinfo {author}
  {\bibfnamefont {K.}~\bibnamefont {Tao}}, \bibinfo {author} {\bibfnamefont
  {C.}~\bibnamefont {Kuperwasser}}, \ and\ \bibinfo {author} {\bibfnamefont
  {E.~S.}\ \bibnamefont {Lander}},\ }\bibfield  {title} {\enquote {\bibinfo
  {title} {{Stochastic State Transitions Give Rise to Phenotypic Equilibrium in
  Populations of Cancer Cells}},}\ }\href@noop {} {\bibfield  {journal}
  {\bibinfo  {journal} {Cell}\ }\textbf {\bibinfo {volume} {146}},\ \bibinfo
  {pages} {633} (\bibinfo {year} {2011})}\BibitemShut {NoStop}%
\bibitem [{\citenamefont {\textit{et} al.}(2011)}]{Chaffer:2011iv}%
  \BibitemOpen
  \bibfield  {author} {\bibinfo {author} {\bibfnamefont {C.~L.~Chaffer}\
  \bibnamefont {\textit{et} al.}},\ }\bibfield  {title} {\enquote {\bibinfo
  {title} {{Normal and Neoplastic Nonstem Cells Can Spontaneously Convert to a
  Stem-Like State}},}\ }\href@noop {} {\bibfield  {journal} {\bibinfo
  {journal} {Proc. Natl. Acad. Sci. U.S.A.}\ }\textbf {\bibinfo {volume}
  {108}},\ \bibinfo {pages} {7950} (\bibinfo {year} {2011})}\BibitemShut
  {NoStop}%
\bibitem [{\citenamefont {Waddington}(1957)}]{WADDINGTON:1957wn}%
  \BibitemOpen
  \bibfield  {author} {\bibinfo {author} {\bibfnamefont {C.~H.}\ \bibnamefont
  {Waddington}},\ }\href@noop {} {\emph {\bibinfo {title} {{The Strategy of the
  Genes}}}}\ (\bibinfo  {publisher} {Allen and Unwin, London},\ \bibinfo {year}
  {1957})\BibitemShut {NoStop}%
\bibitem [{\citenamefont {Huang}\ \emph {et~al.}(2005)\citenamefont {Huang},
  \citenamefont {Eichler}, \citenamefont {Bar-Yam},\ and\ \citenamefont
  {Ingber}}]{Huang:2005fv}%
  \BibitemOpen
  \bibfield  {author} {\bibinfo {author} {\bibfnamefont {S.}~\bibnamefont
  {Huang}}, \bibinfo {author} {\bibfnamefont {G.}~\bibnamefont {Eichler}},
  \bibinfo {author} {\bibfnamefont {Y.}~\bibnamefont {Bar-Yam}}, \ and\
  \bibinfo {author} {\bibfnamefont {D.~E.}\ \bibnamefont {Ingber}},\ }\bibfield
   {title} {\enquote {\bibinfo {title} {{Cell Fates as High-Dimensional
  Attractor States of a Complex Gene Regulatory network}},}\ }\href@noop {}
  {\bibfield  {journal} {\bibinfo  {journal} {Phys. Rev. Lett.}\ }\textbf
  {\bibinfo {volume} {94}},\ \bibinfo {pages} {128701} (\bibinfo {year}
  {2005})}\BibitemShut {NoStop}%
\bibitem [{\citenamefont {Acar}\ \emph {et~al.}(2005)\citenamefont {Acar},
  \citenamefont {Becskei},\ and\ \citenamefont {van
  Oudenaarden}}]{Acar:2005ds}%
  \BibitemOpen
  \bibfield  {author} {\bibinfo {author} {\bibfnamefont {M.}~\bibnamefont
  {Acar}}, \bibinfo {author} {\bibfnamefont {A.}~\bibnamefont {Becskei}}, \
  and\ \bibinfo {author} {\bibfnamefont {A.}~\bibnamefont {van Oudenaarden}},\
  }\bibfield  {title} {\enquote {\bibinfo {title} {{Enhancement of Cellular
  Memory by Reducing Stochastic Transitions}},}\ }\href@noop {} {\bibfield
  {journal} {\bibinfo  {journal} {Nature (London)}\ }\textbf {\bibinfo {volume}
  {435}},\ \bibinfo {pages} {228} (\bibinfo {year} {2005})}\BibitemShut
  {NoStop}%
\bibitem [{\citenamefont {Ishimatsu}\ \emph {et~al.}(2014)\citenamefont
  {Ishimatsu}, \citenamefont {Hata}, \citenamefont {Mochizuki}, \citenamefont
  {Sekine}, \citenamefont {Yamamura},\ and\ \citenamefont {Kiga}}]{Kiga}%
  \BibitemOpen
  \bibfield  {author} {\bibinfo {author} {\bibfnamefont {K.}~\bibnamefont
  {Ishimatsu}}, \bibinfo {author} {\bibfnamefont {T.}~\bibnamefont {Hata}},
  \bibinfo {author} {\bibfnamefont {A.}~\bibnamefont {Mochizuki}}, \bibinfo
  {author} {\bibfnamefont {R.}~\bibnamefont {Sekine}}, \bibinfo {author}
  {\bibfnamefont {M.}~\bibnamefont {Yamamura}}, \ and\ \bibinfo {author}
  {\bibfnamefont {D.}~\bibnamefont {Kiga}},\ }\bibfield  {title} {\enquote
  {\bibinfo {title} {{General Applicability of Synthetic Gene-Overexpresssion
  for Cell-Type Ratio Control via Reprogramming}},}\ }\href@noop {} {\bibfield
  {journal} {\bibinfo  {journal} {ACS. Synth. Biol.}\ }\textbf {\bibinfo
  {volume} {3}},\ \bibinfo {pages} {638} (\bibinfo {year} {2014})}\BibitemShut
  {NoStop}%
\bibitem [{\citenamefont {Dai}\ \emph {et~al.}(2012)\citenamefont {Dai},
  \citenamefont {Vorselen}, \citenamefont {Korolev},\ and\ \citenamefont
  {Gore}}]{Gore}%
  \BibitemOpen
  \bibfield  {author} {\bibinfo {author} {\bibfnamefont {L.}~\bibnamefont
  {Dai}}, \bibinfo {author} {\bibfnamefont {D.}~\bibnamefont {Vorselen}},
  \bibinfo {author} {\bibfnamefont {K.}~\bibnamefont {Korolev}}, \ and\
  \bibinfo {author} {\bibfnamefont {J.}~\bibnamefont {Gore}},\ }\bibfield
  {title} {\enquote {\bibinfo {title} {{Generic Indicators for Loss of
  Resilience Before a Tipping Point Leading to Population Collapse}},}\
  }\href@noop {} {\bibfield  {journal} {\bibinfo  {journal} {Science}\ }\textbf
  {\bibinfo {volume} {336}},\ \bibinfo {pages} {1175} (\bibinfo {year}
  {2012})}\BibitemShut {NoStop}%
\bibitem [{\citenamefont {Hanna}\ \emph {et~al.}(2009)\citenamefont {Hanna},
  \citenamefont {Saha}, \citenamefont {Pando}, \citenamefont {van Zon},
  \citenamefont {Lengner}, \citenamefont {Creyghton}, \citenamefont {van
  Oudenaarden},\ and\ \citenamefont {Jaenisch}}]{Hanna:2009ix}%
  \BibitemOpen
  \bibfield  {author} {\bibinfo {author} {\bibfnamefont {J.}~\bibnamefont
  {Hanna}}, \bibinfo {author} {\bibfnamefont {K.}~\bibnamefont {Saha}},
  \bibinfo {author} {\bibfnamefont {B.}~\bibnamefont {Pando}}, \bibinfo
  {author} {\bibfnamefont {J.}~\bibnamefont {van Zon}}, \bibinfo {author}
  {\bibfnamefont {C.~J.}\ \bibnamefont {Lengner}}, \bibinfo {author}
  {\bibfnamefont {M.~P.}\ \bibnamefont {Creyghton}}, \bibinfo {author}
  {\bibfnamefont {A.}~\bibnamefont {van Oudenaarden}}, \ and\ \bibinfo {author}
  {\bibfnamefont {R.}~\bibnamefont {Jaenisch}},\ }\bibfield  {title} {\enquote
  {\bibinfo {title} {{Direct Cell Reprogramming is a Stochastic Process
  Amenable to Acceleration}},}\ }\href@noop {} {\bibfield  {journal} {\bibinfo
  {journal} {Nature (London)}\ }\textbf {\bibinfo {volume} {462}},\ \bibinfo
  {pages} {595} (\bibinfo {year} {2009})}\BibitemShut {NoStop}%
\bibitem [{\citenamefont {Robinton}\ and\ \citenamefont
  {Daley}(2012)}]{Robinton:2012}%
  \BibitemOpen
  \bibfield  {author} {\bibinfo {author} {\bibfnamefont {D.~A.}\ \bibnamefont
  {Robinton}}\ and\ \bibinfo {author} {\bibfnamefont {G.~Q.}\ \bibnamefont
  {Daley}},\ }\bibfield  {title} {\enquote {\bibinfo {title} {{The Promise of
  Induced Pluripotent Stem Cells in Research and Therapy}},}\ }\href@noop {}
  {\bibfield  {journal} {\bibinfo  {journal} {Nature (London)}\ }\textbf
  {\bibinfo {volume} {481}},\ \bibinfo {pages} {295} (\bibinfo {year}
  {2012})}\BibitemShut {NoStop}%
\bibitem [{\citenamefont {Huang}\ and\ \citenamefont
  {Kauffman}(2013)}]{Huang:2009kj}%
  \BibitemOpen
  \bibfield  {author} {\bibinfo {author} {\bibfnamefont {S.}~\bibnamefont
  {Huang}}\ and\ \bibinfo {author} {\bibfnamefont {S.}~\bibnamefont
  {Kauffman}},\ }\bibfield  {title} {\enquote {\bibinfo {title} {{How to Escape
  the Cancer Attractor: Rationale and Limitations of Multi-Target Drugs}},}\
  }\href@noop {} {\bibfield  {journal} {\bibinfo  {journal} {Semin. Cancer.
  Biol.}\ }\textbf {\bibinfo {volume} {23}},\ \bibinfo {pages} {270} (\bibinfo
  {year} {2013})}\BibitemShut {NoStop}%
\bibitem [{\citenamefont {Creixell}\ \emph {et~al.}(2012)\citenamefont
  {Creixell}, \citenamefont {Schoof}, \citenamefont {Erler},\ and\
  \citenamefont {Linding}}]{Creixell2012}%
  \BibitemOpen
  \bibfield  {author} {\bibinfo {author} {\bibfnamefont {P.}~\bibnamefont
  {Creixell}}, \bibinfo {author} {\bibfnamefont {E.~M.}\ \bibnamefont
  {Schoof}}, \bibinfo {author} {\bibfnamefont {J.~T.}\ \bibnamefont {Erler}}, \
  and\ \bibinfo {author} {\bibfnamefont {R.}~\bibnamefont {Linding}},\
  }\bibfield  {title} {\enquote {\bibinfo {title} {{Navigating Cancer Network
  Attractors for Tumor-Specific Therapy}},}\ }\href@noop {} {\bibfield
  {journal} {\bibinfo  {journal} {Nat. Biotech.}\ }\textbf {\bibinfo {volume}
  {30}},\ \bibinfo {pages} {842} (\bibinfo {year} {2012})}\BibitemShut
  {NoStop}%
\bibitem [{\citenamefont {Motter}\ \emph {et~al.}(2013)\citenamefont {Motter},
  \citenamefont {Myers}, \citenamefont {Anghel},\ and\ \citenamefont
  {Nishikawa}}]{MotterNatPhys}%
  \BibitemOpen
  \bibfield  {author} {\bibinfo {author} {\bibfnamefont {A.~E.}\ \bibnamefont
  {Motter}}, \bibinfo {author} {\bibfnamefont {S.~A.}\ \bibnamefont {Myers}},
  \bibinfo {author} {\bibfnamefont {M.}~\bibnamefont {Anghel}}, \ and\ \bibinfo
  {author} {\bibfnamefont {T.}~\bibnamefont {Nishikawa}},\ }\bibfield  {title}
  {\enquote {\bibinfo {title} {{Spontaneous Synchrony in Power-Grid
  Networks}},}\ }\href@noop {} {\bibfield  {journal} {\bibinfo  {journal} {Nat.
  Phys.}\ }\textbf {\bibinfo {volume} {9}},\ \bibinfo {pages} {191} (\bibinfo
  {year} {2013})}\BibitemShut {NoStop}%
\bibitem [{\citenamefont {Schnurr}\ \emph {et~al.}(2002)\citenamefont
  {Schnurr}, \citenamefont {Gittes},\ and\ \citenamefont
  {MacKintosh}}]{MacKintoshMetastable}%
  \BibitemOpen
  \bibfield  {author} {\bibinfo {author} {\bibfnamefont {B.}~\bibnamefont
  {Schnurr}}, \bibinfo {author} {\bibfnamefont {F.}~\bibnamefont {Gittes}}, \
  and\ \bibinfo {author} {\bibfnamefont {F.~C.}\ \bibnamefont {MacKintosh}},\
  }\bibfield  {title} {\enquote {\bibinfo {title} {{Metastable Intermediates in
  the Condensation of Semiflexible Polymers}},}\ }\href@noop {} {\bibfield
  {journal} {\bibinfo  {journal} {Phys. Rev. E}\ }\textbf {\bibinfo {volume}
  {65}},\ \bibinfo {pages} {061904} (\bibinfo {year} {2002})}\BibitemShut
  {NoStop}%
\bibitem [{\citenamefont {Sahasrabudhe}\ and\ \citenamefont
  {Motter}(2011)}]{Motter2}%
  \BibitemOpen
  \bibfield  {author} {\bibinfo {author} {\bibfnamefont {S.}~\bibnamefont
  {Sahasrabudhe}}\ and\ \bibinfo {author} {\bibfnamefont {A.~E.}\ \bibnamefont
  {Motter}},\ }\bibfield  {title} {\enquote {\bibinfo {title} {{Rescusing
  Ecosystems from Extinction Cascades Through Compensatory Perturbations}},}\
  }\href@noop {} {\bibfield  {journal} {\bibinfo  {journal} {Nat. Commun.}\
  }\textbf {\bibinfo {volume} {2}},\ \bibinfo {pages} {170} (\bibinfo {year}
  {2011})}\BibitemShut {NoStop}%
\bibitem [{Sup()}]{SupplementalMaterial}%
  \BibitemOpen
  \href@noop {} {\bibinfo  {journal} {See Supplemental Material for
  generalizations of OLAC, additional analyses, supplemental table,
  supplemental figures, and supplemental references.}\ }\BibitemShut {NoStop}%
\bibitem [{\citenamefont {Freidlin}\ and\ \citenamefont
  {Wentzell}(2013)}]{Freidlin:2012vi}%
  \BibitemOpen
\bibfield  {journal} {  }\bibfield  {author} {\bibinfo {author} {\bibfnamefont
  {M.~I.}\ \bibnamefont {Freidlin}}\ and\ \bibinfo {author} {\bibfnamefont
  {A.~D.}\ \bibnamefont {Wentzell}},\ }\href@noop {} {\emph {\bibinfo {title}
  {{Random Perturbations of Dynamical Systems}}}}\ (\bibinfo  {publisher}
  {Springer, Berlin},\ \bibinfo {year} {2013})\BibitemShut {NoStop}%
\bibitem [{\citenamefont {Zhou}\ \emph {et~al.}(2008)\citenamefont {Zhou},
  \citenamefont {Ren},\ and\ \citenamefont {E}}]{Zhou:2008jn}%
  \BibitemOpen
  \bibfield  {author} {\bibinfo {author} {\bibfnamefont {X.}~\bibnamefont
  {Zhou}}, \bibinfo {author} {\bibfnamefont {W.}~\bibnamefont {Ren}}, \ and\
  \bibinfo {author} {\bibfnamefont {W.}~\bibnamefont {E}},\ }\bibfield  {title}
  {\enquote {\bibinfo {title} {{Adaptive Minimum Action Method for the Study of
  Rare Events}},}\ }\href@noop {} {\bibfield  {journal} {\bibinfo  {journal}
  {J. Chem. Phys.}\ }\textbf {\bibinfo {volume} {128}},\ \bibinfo {pages}
  {104111} (\bibinfo {year} {2008})}\BibitemShut {NoStop}%
\bibitem [{\citenamefont {Maier}\ and\ \citenamefont
  {Stein}(1997)}]{Maier:1997vx}%
  \BibitemOpen
  \bibfield  {author} {\bibinfo {author} {\bibfnamefont {R.}~\bibnamefont
  {Maier}}\ and\ \bibinfo {author} {\bibfnamefont {D.}~\bibnamefont {Stein}},\
  }\bibfield  {title} {\enquote {\bibinfo {title} {{Limiting Exit Location
  Distributions in the Stochastic Exit Problem}},}\ }\href@noop {} {\bibfield
  {journal} {\bibinfo  {journal} {SIAM J. Appl. Math.}\ }\textbf {\bibinfo
  {volume} {57}},\ \bibinfo {pages} {752} (\bibinfo {year} {1997})}\BibitemShut
  {NoStop}%
\bibitem [{\citenamefont {Corson}\ and\ \citenamefont
  {Siggia}(2012)}]{Corson:2012wm}%
  \BibitemOpen
  \bibfield  {author} {\bibinfo {author} {\bibfnamefont {F.}~\bibnamefont
  {Corson}}\ and\ \bibinfo {author} {\bibfnamefont {E.~D.}\ \bibnamefont
  {Siggia}},\ }\bibfield  {title} {\enquote {\bibinfo {title} {{Geometry,
  Epistasis, and Developmental Patterning}},}\ }\href@noop {} {\bibfield
  {journal} {\bibinfo  {journal} {Proc. Natl. Acad. Sci. U.S.A.}\ }\textbf
  {\bibinfo {volume} {109}},\ \bibinfo {pages} {5568} (\bibinfo {year}
  {2012})}\BibitemShut {NoStop}%
\bibitem [{\citenamefont {Sternberg}(2005)}]{VD:2005dw}%
  \BibitemOpen
  \bibfield  {author} {\bibinfo {author} {\bibfnamefont {P.~W.}\ \bibnamefont
  {Sternberg}},\ }\bibfield  {title} {\enquote {\bibinfo {title} {{Vulval
  Development}},}\ }\href@noop {} {\bibfield  {journal} {\bibinfo  {journal}
  {WormBook: The online review of C. elegans biology}\ }\textbf {\bibinfo
  {volume} {1}},\ \bibinfo {pages} {10.1895/wormbook.1.6.1,
  http://wormbook.org} (\bibinfo {year} {2005})}\BibitemShut {NoStop}%
\bibitem [{\citenamefont {Nocedal}\ and\ \citenamefont
  {Wright}(2006)}]{Nocedal:2006uv}%
  \BibitemOpen
  \bibfield  {author} {\bibinfo {author} {\bibfnamefont {J.}~\bibnamefont
  {Nocedal}}\ and\ \bibinfo {author} {\bibfnamefont {S.~J.}\ \bibnamefont
  {Wright}},\ }\href@noop {} {\emph {\bibinfo {title} {{Numerical
  Optimization}}}}\ (\bibinfo  {publisher} {Springer, Berlin},\ \bibinfo {year}
  {2006})\BibitemShut {NoStop}%
\bibitem [{\citenamefont {Smelyanskiy}\ and\ \citenamefont
  {Dykman}(1997)}]{Dykman}%
  \BibitemOpen
  \bibfield  {author} {\bibinfo {author} {\bibfnamefont {V.~N.}\ \bibnamefont
  {Smelyanskiy}}\ and\ \bibinfo {author} {\bibfnamefont {M.~I.}\ \bibnamefont
  {Dykman}},\ }\bibfield  {title} {\enquote {\bibinfo {title} {{Optimal Control
  of Large Fluctuations}},}\ }\href@noop {} {\bibfield  {journal} {\bibinfo
  {journal} {Phys. Rev. E}\ }\textbf {\bibinfo {volume} {55}},\ \bibinfo
  {pages} {2516} (\bibinfo {year} {1997})}\BibitemShut {NoStop}%
\bibitem [{\citenamefont {M.~I.~Dykman}\ and\ \citenamefont
  {Hunt}(1994)}]{Dykman6}%
  \BibitemOpen
  \bibfield  {author} {\bibinfo {author} {\bibfnamefont {J.~Ross}\ \bibnamefont
  {M.~I.~Dykman}, \bibfnamefont {E.~Mori}}\ and\ \bibinfo {author}
  {\bibfnamefont {P.~M.}\ \bibnamefont {Hunt}},\ }\bibfield  {title} {\enquote
  {\bibinfo {title} {{Large fluctuations and Optimal Paths in Chemical
  Kinetics}},}\ }\href@noop {} {\bibfield  {journal} {\bibinfo  {journal} {J.
  Chem. Phys.}\ }\textbf {\bibinfo {volume} {100}},\ \bibinfo {pages} {5735}
  (\bibinfo {year} {1994})}\BibitemShut {NoStop}%
\bibitem [{\citenamefont {Pisarchik}\ and\ \citenamefont
  {Feudel}(2014)}]{Pisarchik}%
  \BibitemOpen
  \bibfield  {author} {\bibinfo {author} {\bibfnamefont {A.~N.}\ \bibnamefont
  {Pisarchik}}\ and\ \bibinfo {author} {\bibfnamefont {U.}~\bibnamefont
  {Feudel}},\ }\bibfield  {title} {\enquote {\bibinfo {title} {Control of
  multistability},}\ }\href@noop {} {\bibfield  {journal} {\bibinfo  {journal}
  {Phys. Rep.}\ }\textbf {\bibinfo {volume} {540}},\ \bibinfo {pages} {167}
  (\bibinfo {year} {2014})}\BibitemShut {NoStop}%
\bibitem [{\citenamefont {Vugmeister}\ and\ \citenamefont
  {Rabitz}(1997)}]{Rabitz}%
  \BibitemOpen
  \bibfield  {author} {\bibinfo {author} {\bibfnamefont {B.~E.}\ \bibnamefont
  {Vugmeister}}\ and\ \bibinfo {author} {\bibfnamefont {H.}~\bibnamefont
  {Rabitz}},\ }\bibfield  {title} {\enquote {\bibinfo {title} {Cooperating with
  non-equilibrium fluctuations through their optimal control},}\ }\href@noop {}
  {\bibfield  {journal} {\bibinfo  {journal} {Phys. Rev. E}\ }\textbf {\bibinfo
  {volume} {55}},\ \bibinfo {pages} {2522} (\bibinfo {year}
  {1997})}\BibitemShut {NoStop}%
\bibitem [{\citenamefont {Lindley}\ and\ \citenamefont
  {Schwartz}(2013)}]{Schwartz1}%
  \BibitemOpen
  \bibfield  {author} {\bibinfo {author} {\bibfnamefont {B.~S.}\ \bibnamefont
  {Lindley}}\ and\ \bibinfo {author} {\bibfnamefont {I.~B.}\ \bibnamefont
  {Schwartz}},\ }\bibfield  {title} {\enquote {\bibinfo {title} {{An Iterative
  Action Minimizing Method for Computing Optimal Paths in Stochastic Dynamical
  Systems}},}\ }\href@noop {} {\bibfield  {journal} {\bibinfo  {journal} {Phys.
  D}\ }\textbf {\bibinfo {volume} {255}},\ \bibinfo {pages} {22} (\bibinfo
  {year} {2013})}\BibitemShut {NoStop}%
\bibitem [{\citenamefont {Billings}\ \emph {et~al.}(2010)\citenamefont
  {Billings}, \citenamefont {Schwartz}, \citenamefont {McCrary}, \citenamefont
  {Korotkov},\ and\ \citenamefont {Dykman}}]{Billings}%
  \BibitemOpen
  \bibfield  {author} {\bibinfo {author} {\bibfnamefont {L.}~\bibnamefont
  {Billings}}, \bibinfo {author} {\bibfnamefont {I.~B.}\ \bibnamefont
  {Schwartz}}, \bibinfo {author} {\bibfnamefont {M.}~\bibnamefont {McCrary}},
  \bibinfo {author} {\bibfnamefont {A.~N.}\ \bibnamefont {Korotkov}}, \ and\
  \bibinfo {author} {\bibfnamefont {M.~I.}\ \bibnamefont {Dykman}},\ }\bibfield
   {title} {\enquote {\bibinfo {title} {{Switching Exponent Scaling Near
  Bifurcation Points for Non-Gaussian Noise}},}\ }\href@noop {} {\bibfield
  {journal} {\bibinfo  {journal} {Phys. Rev. Lett.}\ }\textbf {\bibinfo
  {volume} {104}},\ \bibinfo {pages} {140601} (\bibinfo {year}
  {2010})}\BibitemShut {NoStop}%
\bibitem [{\citenamefont {Becker}\ and\ \citenamefont
  {Karplus}(1997)}]{Becker97:dw}%
  \BibitemOpen
  \bibfield  {author} {\bibinfo {author} {\bibfnamefont {O.~M.}\ \bibnamefont
  {Becker}}\ and\ \bibinfo {author} {\bibfnamefont {M.}~\bibnamefont
  {Karplus}},\ }\bibfield  {title} {\enquote {\bibinfo {title} {{The Topology
  of Multidimensional Potential Energy Surfaces: Theory and Application to
  Peptide Structure and Kinetics}},}\ }\href@noop {} {\bibfield  {journal}
  {\bibinfo  {journal} {J. Chem. Phys.}\ }\textbf {\bibinfo {volume} {106}},\
  \bibinfo {pages} {1495} (\bibinfo {year} {1997})}\BibitemShut {NoStop}%
\bibitem [{\citenamefont {Rao}\ and\ \citenamefont {Caflisch}(2004)}]{Rao2004}%
  \BibitemOpen
  \bibfield  {author} {\bibinfo {author} {\bibfnamefont {F.}~\bibnamefont
  {Rao}}\ and\ \bibinfo {author} {\bibfnamefont {A.}~\bibnamefont {Caflisch}},\
  }\bibfield  {title} {\enquote {\bibinfo {title} {{The Protein Folding
  Network}},}\ }\href@noop {} {\bibfield  {journal} {\bibinfo  {journal} {J.
  Mol. Biol.}\ }\textbf {\bibinfo {volume} {342}},\ \bibinfo {pages} {299}
  (\bibinfo {year} {2004})}\BibitemShut {NoStop}%
\bibitem [{\citenamefont {Wang}\ \emph {et~al.}(2012)\citenamefont {Wang},
  \citenamefont {Saadatpour},\ and\ \citenamefont {Albert}}]{Reka}%
  \BibitemOpen
  \bibfield  {author} {\bibinfo {author} {\bibfnamefont {R.~S.}\ \bibnamefont
  {Wang}}, \bibinfo {author} {\bibfnamefont {A.}~\bibnamefont {Saadatpour}}, \
  and\ \bibinfo {author} {\bibfnamefont {R.}~\bibnamefont {Albert}},\
  }\bibfield  {title} {\enquote {\bibinfo {title} {{Boolean Modeling in Systems
  Biology: an Overview of Methodology and Applications}},}\ }\href@noop {}
  {\bibfield  {journal} {\bibinfo  {journal} {Phys. Biol.}\ }\textbf {\bibinfo
  {volume} {9}},\ \bibinfo {pages} {055001} (\bibinfo {year}
  {2012})}\BibitemShut {NoStop}%
\bibitem [{\citenamefont {Noe}\ and\ \citenamefont
  {Fischer}(2008)}]{TransNetworks}%
  \BibitemOpen
  \bibfield  {author} {\bibinfo {author} {\bibfnamefont {F.}~\bibnamefont
  {Noe}}\ and\ \bibinfo {author} {\bibfnamefont {S.}~\bibnamefont {Fischer}},\
  }\bibfield  {title} {\enquote {\bibinfo {title} {{Transition Networks for
  Modeling the Kinetics of Conformational Change in Macromolecules}},}\
  }\href@noop {} {\bibfield  {journal} {\bibinfo  {journal} {Curr. Opin.
  Struct. Biol.}\ }\textbf {\bibinfo {volume} {18}},\ \bibinfo {pages} {154}
  (\bibinfo {year} {2008})}\BibitemShut {NoStop}%
\bibitem [{\citenamefont {Zhou}\ \emph {et~al.}(2011)\citenamefont {Zhou},
  \citenamefont {Brusch},\ and\ \citenamefont {Huang}}]{Zhou:2011ju}%
  \BibitemOpen
  \bibfield  {author} {\bibinfo {author} {\bibfnamefont {J.~X.}\ \bibnamefont
  {Zhou}}, \bibinfo {author} {\bibfnamefont {L.}~\bibnamefont {Brusch}}, \ and\
  \bibinfo {author} {\bibfnamefont {S.}~\bibnamefont {Huang}},\ }\bibfield
  {title} {\enquote {\bibinfo {title} {{Predicting Pancreas Cell Fate Decisions
  and Reprogramming with a Hierarchical Multi-Attractor Model}},}\ }\href@noop
  {} {\bibfield  {journal} {\bibinfo  {journal} {PLOS ONE}\ }\textbf {\bibinfo
  {volume} {6}},\ \bibinfo {pages} {e14752} (\bibinfo {year}
  {2011})}\BibitemShut {NoStop}%
\bibitem [{\citenamefont {Passier}\ and\ \citenamefont
  {Mummery}(2010)}]{Passier:2010dw}%
  \BibitemOpen
  \bibfield  {author} {\bibinfo {author} {\bibfnamefont {R.}~\bibnamefont
  {Passier}}\ and\ \bibinfo {author} {\bibfnamefont {C.}~\bibnamefont
  {Mummery}},\ }\bibfield  {title} {\enquote {\bibinfo {title} {{Getting to the
  Heart of the Matter: Direct Reprogramming to Cardiomyocytes}},}\ }\href@noop
  {} {\bibfield  {journal} {\bibinfo  {journal} {Cell Stem Cell}\ }\textbf
  {\bibinfo {volume} {7}},\ \bibinfo {pages} {139} (\bibinfo {year}
  {2010})}\BibitemShut {NoStop}%
\bibitem [{\citenamefont {Hanahan}\ and\ \citenamefont
  {Weinberg}(2000)}]{hallmarks:2000dw}%
  \BibitemOpen
  \bibfield  {author} {\bibinfo {author} {\bibfnamefont {D.}~\bibnamefont
  {Hanahan}}\ and\ \bibinfo {author} {\bibfnamefont {R.~A.}\ \bibnamefont
  {Weinberg}},\ }\bibfield  {title} {\enquote {\bibinfo {title} {{The Hallmarks
  of Cancer}},}\ }\href@noop {} {\bibfield  {journal} {\bibinfo  {journal}
  {Cell}\ }\textbf {\bibinfo {volume} {100}},\ \bibinfo {pages} {57} (\bibinfo
  {year} {2000})}\BibitemShut {NoStop}%
\bibitem [{\citenamefont {Calzone}\ \emph {et~al.}(2010)\citenamefont
  {Calzone}, \citenamefont {Tournier}, \citenamefont {Fourquet}, \citenamefont
  {Thieffrey}, \citenamefont {Zhivotovsky}, \citenamefont {Barillot},\ and\
  \citenamefont {Zinovyev}}]{Calzone:2010dw}%
  \BibitemOpen
  \bibfield  {author} {\bibinfo {author} {\bibfnamefont {L.}~\bibnamefont
  {Calzone}}, \bibinfo {author} {\bibfnamefont {L.}~\bibnamefont {Tournier}},
  \bibinfo {author} {\bibfnamefont {S.}~\bibnamefont {Fourquet}}, \bibinfo
  {author} {\bibfnamefont {D.}~\bibnamefont {Thieffrey}}, \bibinfo {author}
  {\bibfnamefont {B.}~\bibnamefont {Zhivotovsky}}, \bibinfo {author}
  {\bibfnamefont {E.}~\bibnamefont {Barillot}}, \ and\ \bibinfo {author}
  {\bibfnamefont {A.}~\bibnamefont {Zinovyev}},\ }\bibfield  {title} {\enquote
  {\bibinfo {title} {{Mathematical Modeling of Cell-Fate Decision in Response
  to Death Receptor Engagement}},}\ }\href@noop {} {\bibfield  {journal}
  {\bibinfo  {journal} {PLOS Comput. Biol.}\ }\textbf {\bibinfo {volume} {6}},\
  \bibinfo {pages} {e1000702} (\bibinfo {year} {2010})}\BibitemShut {NoStop}%
\bibitem [{\citenamefont {Wittmann}\ \emph {et~al.}(2009)\citenamefont
  {Wittmann}, \citenamefont {Krumsiek}, \citenamefont {Saez-Rodriguez},
  \citenamefont {Lauffenburger}, \citenamefont {Klamt},\ and\ \citenamefont
  {Theis}}]{Wittman:2009dw}%
  \BibitemOpen
  \bibfield  {author} {\bibinfo {author} {\bibfnamefont {D.~M.}\ \bibnamefont
  {Wittmann}}, \bibinfo {author} {\bibfnamefont {J.}~\bibnamefont {Krumsiek}},
  \bibinfo {author} {\bibfnamefont {J.}~\bibnamefont {Saez-Rodriguez}},
  \bibinfo {author} {\bibfnamefont {D.~A.}\ \bibnamefont {Lauffenburger}},
  \bibinfo {author} {\bibfnamefont {S.}~\bibnamefont {Klamt}}, \ and\ \bibinfo
  {author} {\bibfnamefont {F.~J.}\ \bibnamefont {Theis}},\ }\bibfield  {title}
  {\enquote {\bibinfo {title} {{Transforming Boolean Models to Continuous
  Models: Methodology and Application to T-cell Receptor Signaling}},}\
  }\href@noop {} {\bibfield  {journal} {\bibinfo  {journal} {BMC Sys. Biol.}\
  }\textbf {\bibinfo {volume} {3}},\ \bibinfo {pages} {98} (\bibinfo {year}
  {2009})}\BibitemShut {NoStop}%
\bibitem [{\citenamefont {Luo}\ \emph {et~al.}(2005)\citenamefont {Luo},
  \citenamefont {Kamata},\ and\ \citenamefont {Karin}}]{Luo:2005dw}%
  \BibitemOpen
  \bibfield  {author} {\bibinfo {author} {\bibfnamefont {J.~L.}\ \bibnamefont
  {Luo}}, \bibinfo {author} {\bibfnamefont {H.}~\bibnamefont {Kamata}}, \ and\
  \bibinfo {author} {\bibfnamefont {M.}~\bibnamefont {Karin}},\ }\bibfield
  {title} {\enquote {\bibinfo {title} {{IKK/NF-$\kappa$B Signaling: Balancing
  Life and Death---a New Approach to Cancer Therapy}},}\ }\href@noop {}
  {\bibfield  {journal} {\bibinfo  {journal} {J. Clin. Invest.}\ }\textbf
  {\bibinfo {volume} {115}},\ \bibinfo {pages} {2625} (\bibinfo {year}
  {2005})}\BibitemShut {NoStop}%
\bibitem [{\citenamefont {Safa}(2012)}]{Safa:dw2012}%
  \BibitemOpen
  \bibfield  {author} {\bibinfo {author} {\bibfnamefont {A.~R.}\ \bibnamefont
  {Safa}},\ }\bibfield  {title} {\enquote {\bibinfo {title} {{c-Flip, a Master
  Anti-Apoptotic Regulator}},}\ }\href@noop {} {\bibfield  {journal} {\bibinfo
  {journal} {Exp. Oncol.}\ }\textbf {\bibinfo {volume} {34}},\ \bibinfo {pages}
  {176} (\bibinfo {year} {2012})}\BibitemShut {NoStop}%
\bibitem [{\citenamefont {Saadatpour}\ \emph {et~al.}(2011)\citenamefont
  {Saadatpour}, \citenamefont {Wang}, \citenamefont {Liao}, \citenamefont
  {Liu}, \citenamefont {Loughran}, \citenamefont {Albert},\ and\ \citenamefont
  {Albert}}]{Saadatpour2011}%
  \BibitemOpen
  \bibfield  {author} {\bibinfo {author} {\bibfnamefont {A.}~\bibnamefont
  {Saadatpour}}, \bibinfo {author} {\bibfnamefont {R.}~\bibnamefont {Wang}},
  \bibinfo {author} {\bibfnamefont {A.}~\bibnamefont {Liao}}, \bibinfo {author}
  {\bibfnamefont {X.}~\bibnamefont {Liu}}, \bibinfo {author} {\bibfnamefont
  {T.~P}\ \bibnamefont {Loughran}}, \bibinfo {author} {\bibfnamefont
  {I.}~\bibnamefont {Albert}}, \ and\ \bibinfo {author} {\bibfnamefont
  {R.}~\bibnamefont {Albert}},\ }\bibfield  {title} {\enquote {\bibinfo {title}
  {Dynamical and structural analysis of a t-cell survival network identifies
  novel candidate therapeutic targets for large granular lymphocyte
  leukemia},}\ }\href@noop {} {\bibfield  {journal} {\bibinfo  {journal} {PLOS
  Comput. Biol.}\ }\textbf {\bibinfo {volume} {7}},\ \bibinfo {pages}
  {e1002267} (\bibinfo {year} {2011})}\BibitemShut {NoStop}%
\bibitem [{\citenamefont {Regan}\ and\ \citenamefont {Aird}(2012)}]{Regan2012}%
  \BibitemOpen
  \bibfield  {author} {\bibinfo {author} {\bibfnamefont {E.~R.}\ \bibnamefont
  {Regan}}\ and\ \bibinfo {author} {\bibfnamefont {W.~C.}\ \bibnamefont
  {Aird}},\ }\bibfield  {title} {\enquote {\bibinfo {title} {{Dynamical Systems
  Approach to Endothelial Heterogeneity}},}\ }\href@noop {} {\bibfield
  {journal} {\bibinfo  {journal} {Circ. Res.}\ }\textbf {\bibinfo {volume}
  {111}},\ \bibinfo {pages} {110} (\bibinfo {year} {2012})}\BibitemShut
  {NoStop}%
\bibitem [{\citenamefont {Lang}\ \emph {et~al.}(2014)\citenamefont {Lang},
  \citenamefont {Li}, \citenamefont {Collins},\ and\ \citenamefont
  {Mehta}}]{Lang}%
  \BibitemOpen
  \bibfield  {author} {\bibinfo {author} {\bibfnamefont {A.~H.}\ \bibnamefont
  {Lang}}, \bibinfo {author} {\bibfnamefont {H.}~\bibnamefont {Li}}, \bibinfo
  {author} {\bibfnamefont {J.~J.}\ \bibnamefont {Collins}}, \ and\ \bibinfo
  {author} {\bibfnamefont {P.}~\bibnamefont {Mehta}},\ }\bibfield  {title}
  {\enquote {\bibinfo {title} {{Epigenetic Landscapes Explain Partially
  Reprogrammed Cells and Identify Key Reprogramming Genes}},}\ }\href@noop {}
  {\bibfield  {journal} {\bibinfo  {journal} {PLOS Comput. Biol.}\ }\textbf
  {\bibinfo {volume} {10}},\ \bibinfo {pages} {e1003734} (\bibinfo {year}
  {2014})}\BibitemShut {NoStop}%
\bibitem [{\citenamefont {Steinway}\ \emph {et~al.}(2014)\citenamefont
  {Steinway}, \citenamefont {Za{\~n}udo}, \citenamefont {Ding}, \citenamefont
  {B.}, \citenamefont {Rountree}, \citenamefont {Feith}, \citenamefont {Jr.},\
  and\ \citenamefont {Albert}}]{Reka3}%
  \BibitemOpen
  \bibfield  {author} {\bibinfo {author} {\bibfnamefont {S.~N.}\ \bibnamefont
  {Steinway}}, \bibinfo {author} {\bibfnamefont {J.~G.~T.}\ \bibnamefont
  {Za{\~n}udo}}, \bibinfo {author} {\bibfnamefont {W.}~\bibnamefont {Ding}},
  \bibinfo {author} {\bibfnamefont {C.}~\bibnamefont {B.}}, \bibinfo {author}
  {\bibnamefont {Rountree}}, \bibinfo {author} {\bibfnamefont {D.~J.}\
  \bibnamefont {Feith}}, \bibinfo {author} {\bibfnamefont {T.~P.~Loughran}\
  \bibnamefont {Jr.}}, \ and\ \bibinfo {author} {\bibfnamefont
  {R.}~\bibnamefont {Albert}},\ }\bibfield  {title} {\enquote {\bibinfo {title}
  {{Network Modeling of TGF$\beta$ Signaling in Hepatocellular Carcinoma
  Epithelial-to-Mesenchymal Transition Reveals Joint Sonic Hedgehog and Wnt
  Pathway Activation}},}\ }\href@noop {} {\bibfield  {journal} {\bibinfo
  {journal} {Cancer Res.}\ }\textbf {\bibinfo {volume} {74}},\ \bibinfo {pages}
  {5963} (\bibinfo {year} {2014})}\BibitemShut {NoStop}%
\bibitem [{\citenamefont {Rais}\ \emph {et~al.}(2013)\citenamefont {Rais},
  \citenamefont {Zviran}, \citenamefont {Geula}, \citenamefont {Gafni},
  \citenamefont {Chomsky}, \citenamefont {Viukov}, \citenamefont {Mansour},
  \citenamefont {Caspi}, \citenamefont {Krupalnik},\ and\ \citenamefont {et.
  al.}}]{Hanna2}%
  \BibitemOpen
  \bibfield  {author} {\bibinfo {author} {\bibfnamefont {Y.}~\bibnamefont
  {Rais}}, \bibinfo {author} {\bibfnamefont {A.}~\bibnamefont {Zviran}},
  \bibinfo {author} {\bibfnamefont {S.}~\bibnamefont {Geula}}, \bibinfo
  {author} {\bibfnamefont {O.}~\bibnamefont {Gafni}}, \bibinfo {author}
  {\bibfnamefont {E.}~\bibnamefont {Chomsky}}, \bibinfo {author} {\bibfnamefont
  {S.}~\bibnamefont {Viukov}}, \bibinfo {author} {\bibfnamefont {A.~AlFatah}\
  \bibnamefont {Mansour}}, \bibinfo {author} {\bibfnamefont {I.}~\bibnamefont
  {Caspi}}, \bibinfo {author} {\bibfnamefont {V.}~\bibnamefont {Krupalnik}}, \
  and\ \bibinfo {author} {\bibfnamefont {M.~Zerbib}\ \bibnamefont {et. al.}},\
  }\bibfield  {title} {\enquote {\bibinfo {title} {{Deterministic Direct
  Reprogramming of Somatic Cells to Pluripotency}},}\ }\href@noop {} {\bibfield
   {journal} {\bibinfo  {journal} {Nature (London)}\ }\textbf {\bibinfo
  {volume} {502}},\ \bibinfo {pages} {65} (\bibinfo {year} {2013})}\BibitemShut
  {NoStop}%
\bibitem [{\citenamefont {Turner}\ \emph {et~al.}(2004)\citenamefont {Turner},
  \citenamefont {Schnell},\ and\ \citenamefont {Burrage}}]{StochRev}%
  \BibitemOpen
  \bibfield  {author} {\bibinfo {author} {\bibfnamefont {T.~E.}\ \bibnamefont
  {Turner}}, \bibinfo {author} {\bibfnamefont {S.}~\bibnamefont {Schnell}}, \
  and\ \bibinfo {author} {\bibfnamefont {K.}~\bibnamefont {Burrage}},\
  }\bibfield  {title} {\enquote {\bibinfo {title} {{Stochastic Approaches for
  Modelling In Vivo Reactions}},}\ }\href@noop {} {\bibfield  {journal}
  {\bibinfo  {journal} {Comput. Biol. Chem.}\ }\textbf {\bibinfo {volume}
  {18}},\ \bibinfo {pages} {165} (\bibinfo {year} {2004})}\BibitemShut
  {NoStop}%
\bibitem [{\citenamefont {Liu}(2006)}]{DiLiu1}%
  \BibitemOpen
  \bibfield  {author} {\bibinfo {author} {\bibfnamefont {D.}~\bibnamefont
  {Liu}},\ }\bibfield  {title} {\enquote {\bibinfo {title} {{Optimal Transition
  Paths of Stochastic Chemical Kinetic Systems}},}\ }\href@noop {} {\bibfield
  {journal} {\bibinfo  {journal} {J. Chem. Phys.}\ }\textbf {\bibinfo {volume}
  {124}},\ \bibinfo {pages} {164104} (\bibinfo {year} {2006})}\BibitemShut
  {NoStop}%
\bibitem [{\citenamefont {Liu}(2008)}]{DiLiu2}%
  \BibitemOpen
  \bibfield  {author} {\bibinfo {author} {\bibfnamefont {D.}~\bibnamefont
  {Liu}},\ }\bibfield  {title} {\enquote {\bibinfo {title} {{A Numerical Scheme
  for Optimal Transition Paths of Stochastic Chemical Kinetic Systems}},}\
  }\href@noop {} {\bibfield  {journal} {\bibinfo  {journal} {J. Comp. Phys}\
  }\textbf {\bibinfo {volume} {227}},\ \bibinfo {pages} {8672} (\bibinfo {year}
  {2008})}\BibitemShut {NoStop}%
\bibitem [{\citenamefont {Gaj}\ \emph {et~al.}(2013)\citenamefont {Gaj},
  \citenamefont {Gersbach},\ and\ \citenamefont {III}}]{GenomeEditing}%
  \BibitemOpen
  \bibfield  {author} {\bibinfo {author} {\bibfnamefont {T.}~\bibnamefont
  {Gaj}}, \bibinfo {author} {\bibfnamefont {C.~A.}\ \bibnamefont {Gersbach}}, \
  and\ \bibinfo {author} {\bibfnamefont {C.~F.~Barbas}\ \bibnamefont {III}},\
  }\bibfield  {title} {\enquote {\bibinfo {title} {{ZFN, TALEN, and
  CRISPR/Cas-Based Methods for Genome Engineering}},}\ }\href@noop {}
  {\bibfield  {journal} {\bibinfo  {journal} {Trends in Biotechnol.}\ }\textbf
  {\bibinfo {volume} {31}},\ \bibinfo {pages} {397} (\bibinfo {year}
  {2013})}\BibitemShut {NoStop}%
\bibitem [{\citenamefont {Cornelius}\ \emph {et~al.}(2013)\citenamefont
  {Cornelius}, \citenamefont {Kath},\ and\ \citenamefont {Motter}}]{Sean}%
  \BibitemOpen
  \bibfield  {author} {\bibinfo {author} {\bibfnamefont {S.~P.}\ \bibnamefont
  {Cornelius}}, \bibinfo {author} {\bibfnamefont {W.~L.}\ \bibnamefont {Kath}},
  \ and\ \bibinfo {author} {\bibfnamefont {A.~E.}\ \bibnamefont {Motter}},\
  }\bibfield  {title} {\enquote {\bibinfo {title} {{Realistic Control of
  Network Dynamics}},}\ }\href@noop {} {\bibfield  {journal} {\bibinfo
  {journal} {Nat. Commun.}\ }\textbf {\bibinfo {volume} {4}},\ \bibinfo {pages}
  {1942} (\bibinfo {year} {2013})}\BibitemShut {NoStop}%
\bibitem [{\citenamefont {Xia}\ \emph {et~al.}(2002)\citenamefont {Xia},
  \citenamefont {Mao}, \citenamefont {Paulson},\ and\ \citenamefont
  {Davidson}}]{RealsiRNA}%
  \BibitemOpen
  \bibfield  {author} {\bibinfo {author} {\bibfnamefont {H.}~\bibnamefont
  {Xia}}, \bibinfo {author} {\bibfnamefont {Q.}~\bibnamefont {Mao}}, \bibinfo
  {author} {\bibfnamefont {H.}~\bibnamefont {Paulson}}, \ and\ \bibinfo
  {author} {\bibfnamefont {B.~L.}\ \bibnamefont {Davidson}},\ }\bibfield
  {title} {\enquote {\bibinfo {title} {{siRNA-Mediated Gene Silencing in vitro
  and in vivo}},}\ }\href@noop {} {\bibfield  {journal} {\bibinfo  {journal}
  {Nat. Biotechnol.}\ }\textbf {\bibinfo {volume} {20}},\ \bibinfo {pages}
  {1006} (\bibinfo {year} {2002})}\BibitemShut {NoStop}%
\bibitem [{\citenamefont {Za{\~n}udo}\ and\ \citenamefont
  {Albert}(2015)}]{AlbertArx}%
  \BibitemOpen
  \bibfield  {author} {\bibinfo {author} {\bibfnamefont {J.~G.~T.}\
  \bibnamefont {Za{\~n}udo}}\ and\ \bibinfo {author} {\bibfnamefont
  {R.}~\bibnamefont {Albert}},\ }\bibfield  {title} {\enquote {\bibinfo {title}
  {{Cell Fate Reprogramming by Control of Intracellular Network Dynamics}},}\
  }\href@noop {} {\bibfield  {journal} {\bibinfo  {journal} {PLOS Comput.
  Biol.}\ }\textbf {\bibinfo {volume} {11}},\ \bibinfo {pages} {e1004193}
  (\bibinfo {year} {2015})}\BibitemShut {NoStop}%
\bibitem [{\citenamefont {Wang}\ \emph {et~al.}(2011)\citenamefont {Wang},
  \citenamefont {Zhang}, \citenamefont {Xu},\ and\ \citenamefont
  {Wang}}]{Wang:2011ef}%
  \BibitemOpen
  \bibfield  {author} {\bibinfo {author} {\bibfnamefont {J.}~\bibnamefont
  {Wang}}, \bibinfo {author} {\bibfnamefont {K.}~\bibnamefont {Zhang}},
  \bibinfo {author} {\bibfnamefont {L.}~\bibnamefont {Xu}}, \ and\ \bibinfo
  {author} {\bibfnamefont {E.}~\bibnamefont {Wang}},\ }\bibfield  {title}
  {\enquote {\bibinfo {title} {{Quantifying the Waddington Landscape and
  Biological Paths for Development and Differentiation}},}\ }\href@noop {}
  {\bibfield  {journal} {\bibinfo  {journal} {Proc. Natl. Acad. Sci. U.S.A.}\
  }\textbf {\bibinfo {volume} {108}},\ \bibinfo {pages} {8257} (\bibinfo {year}
  {2011})}\BibitemShut {NoStop}%
\bibitem [{\citenamefont {Chickarmane}\ and\ \citenamefont
  {Peterson}(2008)}]{Chickarmane:2008ha}%
  \BibitemOpen
  \bibfield  {author} {\bibinfo {author} {\bibfnamefont {V.}~\bibnamefont
  {Chickarmane}}\ and\ \bibinfo {author} {\bibfnamefont {C.}~\bibnamefont
  {Peterson}},\ }\bibfield  {title} {\enquote {\bibinfo {title} {{A
  Computational Model for Understanding Stem Cell, Trophectoderm and Endoderm
  Lineage Determination}},}\ }\href@noop {} {\bibfield  {journal} {\bibinfo
  {journal} {PLOS ONE}\ }\textbf {\bibinfo {volume} {3}},\ \bibinfo {pages}
  {e3478} (\bibinfo {year} {2008})}\BibitemShut {NoStop}%
\bibitem [{\citenamefont {Marchisio}\ and\ \citenamefont
  {Stelling}(2009)}]{Clancy:2010jv}%
  \BibitemOpen
  \bibfield  {author} {\bibinfo {author} {\bibfnamefont {M.}~\bibnamefont
  {Marchisio}}\ and\ \bibinfo {author} {\bibfnamefont {J.}~\bibnamefont
  {Stelling}},\ }\bibfield  {title} {\enquote {\bibinfo {title} {{Computational
  Design Tools for Synthetic Biology}},}\ }\href@noop {} {\bibfield  {journal}
  {\bibinfo  {journal} {Curr. Opin. Biotechnol.}\ }\textbf {\bibinfo {volume}
  {20}},\ \bibinfo {pages} {479} (\bibinfo {year} {2009})}\BibitemShut
  {NoStop}%
\bibitem [{\citenamefont {Menck}\ \emph {et~al.}(2013)\citenamefont {Menck},
  \citenamefont {Heitzig}, \citenamefont {Marwan},\ and\ \citenamefont
  {Kurths}}]{Kurths}%
  \BibitemOpen
  \bibfield  {author} {\bibinfo {author} {\bibfnamefont {P.~J.}\ \bibnamefont
  {Menck}}, \bibinfo {author} {\bibfnamefont {J.}~\bibnamefont {Heitzig}},
  \bibinfo {author} {\bibfnamefont {N.}~\bibnamefont {Marwan}}, \ and\ \bibinfo
  {author} {\bibfnamefont {J.}~\bibnamefont {Kurths}},\ }\bibfield  {title}
  {\enquote {\bibinfo {title} {{How Basin Stability Complements the
  Linear-Stability Paradigm}},}\ }\href@noop {} {\bibfield  {journal} {\bibinfo
   {journal} {Nat. Phys.}\ }\textbf {\bibinfo {volume} {9}},\ \bibinfo {pages}
  {89} (\bibinfo {year} {2013})}\BibitemShut {NoStop}%
\bibitem [{\citenamefont {Feudel}(2008)}]{Feudel2008}%
  \BibitemOpen
  \bibfield  {author} {\bibinfo {author} {\bibfnamefont {U.}~\bibnamefont
  {Feudel}},\ }\bibfield  {title} {\enquote {\bibinfo {title} {{Complex
  Dynamics in Multistable Systems}},}\ }\href@noop {} {\bibfield  {journal}
  {\bibinfo  {journal} {Int. J .Bifurcat. Chaos}\ }\textbf {\bibinfo {volume}
  {18}},\ \bibinfo {pages} {1607} (\bibinfo {year} {2008})}\BibitemShut
  {NoStop}%
\bibitem [{\citenamefont {Granovetter}(1978)}]{Granovetter}%
  \BibitemOpen
  \bibfield  {author} {\bibinfo {author} {\bibfnamefont {M.}~\bibnamefont
  {Granovetter}},\ }\bibfield  {title} {\enquote {\bibinfo {title} {{Threshold
  Models of Collective Behavior}},}\ }\href@noop {} {\bibfield  {journal}
  {\bibinfo  {journal} {Am. J. Sociol.}\ }\textbf {\bibinfo {volume} {83}},\
  \bibinfo {pages} {1420} (\bibinfo {year} {1978})}\BibitemShut {NoStop}%
\bibitem [{\citenamefont {Abrams}\ and\ \citenamefont
  {Strogatz}(2004)}]{Abrams}%
  \BibitemOpen
  \bibfield  {author} {\bibinfo {author} {\bibfnamefont {D.~M.}\ \bibnamefont
  {Abrams}}\ and\ \bibinfo {author} {\bibfnamefont {S.~H.}\ \bibnamefont
  {Strogatz}},\ }\bibfield  {title} {\enquote {\bibinfo {title} {{Chimera
  States for Coupled Oscillators}},}\ }\href@noop {} {\bibfield  {journal}
  {\bibinfo  {journal} {Phys. Rev. Lett.}\ }\textbf {\bibinfo {volume} {93}},\
  \bibinfo {pages} {174102} (\bibinfo {year} {2004})}\BibitemShut {NoStop}%
\bibitem [{\citenamefont {Hopfield}(1982)}]{Hopfield1}%
  \BibitemOpen
  \bibfield  {author} {\bibinfo {author} {\bibfnamefont {J.~J.}\ \bibnamefont
  {Hopfield}},\ }\bibfield  {title} {\enquote {\bibinfo {title} {{Neural
  Networks and Physical Systems with Emergent Computational Abilities}},}\
  }\href@noop {} {\bibfield  {journal} {\bibinfo  {journal} {Proc. Natl. Acad.
  Sci.}\ }\textbf {\bibinfo {volume} {79}},\ \bibinfo {pages} {2554} (\bibinfo
  {year} {1982})}\BibitemShut {NoStop}%
\bibitem [{\citenamefont {Campbell}\ and\ \citenamefont
  {Albert}(2014)}]{Campbell}%
  \BibitemOpen
  \bibfield  {author} {\bibinfo {author} {\bibfnamefont {C.}~\bibnamefont
  {Campbell}}\ and\ \bibinfo {author} {\bibfnamefont {R.}~\bibnamefont
  {Albert}},\ }\bibfield  {title} {\enquote {\bibinfo {title} {{Stabilization
  of Perturbed Boolean Network Attractors Through Compensatory
  Interactions}},}\ }\href@noop {} {\bibfield  {journal} {\bibinfo  {journal}
  {BMC Syst. Biol.}\ }\textbf {\bibinfo {volume} {8}},\ \bibinfo {pages} {53}
  (\bibinfo {year} {2014})}\BibitemShut {NoStop}%
\bibitem [{\citenamefont {M.~Assaf}\ and\ \citenamefont
  {Meerson}(2008)}]{Assaf1}%
  \BibitemOpen
  \bibfield  {author} {\bibinfo {author} {\bibfnamefont {A.~Kamenev}\
  \bibnamefont {M.~Assaf}}\ and\ \bibinfo {author} {\bibfnamefont
  {B.}~\bibnamefont {Meerson}},\ }\bibfield  {title} {\enquote {\bibinfo
  {title} {{Population extinction in a time-modulated environment}},}\
  }\href@noop {} {\bibfield  {journal} {\bibinfo  {journal} {Phys. Rev. E.}\
  }\textbf {\bibinfo {volume} {78}},\ \bibinfo {pages} {041123} (\bibinfo
  {year} {2008})}\BibitemShut {NoStop}%
\bibitem [{\citenamefont {M.~I.~Dykman}\ and\ \citenamefont
  {Landsman}(2008)}]{DykmanSchwartz1}%
  \BibitemOpen
  \bibfield  {author} {\bibinfo {author} {\bibfnamefont {I.~B.~Schwartz}\
  \bibnamefont {M.~I.~Dykman}}\ and\ \bibinfo {author} {\bibfnamefont {A.~S.}\
  \bibnamefont {Landsman}},\ }\bibfield  {title} {\enquote {\bibinfo {title}
  {{Disease Extinction in the Presence of Random Vaccination}},}\ }\href@noop
  {} {\bibfield  {journal} {\bibinfo  {journal} {Phys. Rev. Lett.}\ }\textbf
  {\bibinfo {volume} {101}},\ \bibinfo {pages} {078101} (\bibinfo {year}
  {2008})}\BibitemShut {NoStop}%
\bibitem [{\citenamefont {Khasin}\ and\ \citenamefont
  {Dykman}(2009)}]{Dykman1}%
  \BibitemOpen
  \bibfield  {author} {\bibinfo {author} {\bibfnamefont {M.}~\bibnamefont
  {Khasin}}\ and\ \bibinfo {author} {\bibfnamefont {M.~I.}\ \bibnamefont
  {Dykman}},\ }\bibfield  {title} {\enquote {\bibinfo {title} {{Extinction rate
  fragility in population dynamics}},}\ }\href@noop {} {\bibfield  {journal}
  {\bibinfo  {journal} {Phys. Rev. Lett.}\ }\textbf {\bibinfo {volume} {103}},\
  \bibinfo {pages} {068101} (\bibinfo {year} {2009})}\BibitemShut {NoStop}%
\bibitem [{\citenamefont {C.~R.~Doering}\ and\ \citenamefont
  {Sander}(2005)}]{Doering1}%
  \BibitemOpen
  \bibfield  {author} {\bibinfo {author} {\bibfnamefont {K.~V.~Sargsyan}\
  \bibnamefont {C.~R.~Doering}}\ and\ \bibinfo {author} {\bibfnamefont {L.~M.}\
  \bibnamefont {Sander}},\ }\bibfield  {title} {\enquote {\bibinfo {title}
  {{Extinction times for birth-death processes: exact results, continuum
  asymptotics, and the failure of the Fokker--Planck approximation}},}\
  }\href@noop {} {\bibfield  {journal} {\bibinfo  {journal} {Multiscale Model.
  Simul.}\ }\textbf {\bibinfo {volume} {3}},\ \bibinfo {pages} {283} (\bibinfo
  {year} {2005})}\BibitemShut {NoStop}%
\bibitem [{\citenamefont {Karatzas}\ and\ \citenamefont
  {Shreve}(1991)}]{Karatzas:1991ws}%
  \BibitemOpen
  \bibfield  {author} {\bibinfo {author} {\bibfnamefont {I.}~\bibnamefont
  {Karatzas}}\ and\ \bibinfo {author} {\bibfnamefont {S.~E.}\ \bibnamefont
  {Shreve}},\ }\href@noop {} {\emph {\bibinfo {title} {{Brownian Motion and
  Stochastic Calculus}}}}\ (\bibinfo  {publisher} {Springer, Berlin},\ \bibinfo
  {year} {1991})\BibitemShut {NoStop}%
\bibitem [{\citenamefont {Seydel}(1988)}]{Seydel:1988vz}%
  \BibitemOpen
  \bibfield  {author} {\bibinfo {author} {\bibfnamefont {R.}~\bibnamefont
  {Seydel}},\ }\href@noop {} {\emph {\bibinfo {title} {{From Equilibrium to
  Chaos: Practical Bifurcation and Stability Analysis}}}}\ (\bibinfo
  {publisher} {Elsevier, Amsterdam},\ \bibinfo {year} {1988})\BibitemShut
  {NoStop}%
\bibitem [{2dw(2013)}]{2dw}%
  \BibitemOpen
  \href@noop {} {\emph {\bibinfo {title} {{MATLAB version 8.1.0.604}}}}\
  (\bibinfo  {publisher} {The MathWorks Inc. Natick, Massachusetts},\ \bibinfo
  {year} {2013})\BibitemShut {NoStop}%
\bibitem [{\citenamefont {Boyd}\ and\ \citenamefont
  {Vandenberghe}(2004)}]{6dw}%
  \BibitemOpen
  \bibfield  {author} {\bibinfo {author} {\bibfnamefont {S.}~\bibnamefont
  {Boyd}}\ and\ \bibinfo {author} {\bibfnamefont {L.}~\bibnamefont
  {Vandenberghe}},\ }\href@noop {} {\emph {\bibinfo {title} {{Convex
  Optimization}}}}\ (\bibinfo  {publisher} {Cambridge University Press, New
  York, NY},\ \bibinfo {year} {2004})\BibitemShut {NoStop}%
\end{thebibliography}
